\documentclass[AMA,STIX1COL]{WileyNJD-v2}
\usepackage{url}
\usepackage{moreverb,mathtools,amsfonts,amsmath,amssymb,bbold,float}
\usepackage[algo2e]{algorithm2e} 
\usepackage{algorithm}
\usepackage{algpseudocode}
\usepackage{color}
\usepackage{booktabs} 
\usepackage{xr}
\usepackage{graphicx}
\usepackage{color, soul}                      % Colors
\usepackage{amsmath}
\usepackage{amssymb}
\usepackage{mathtools}
\usepackage{moreverb,mathtools,amsfonts,amssymb,bbold,float}
\usepackage{scalerel,stackengine}       % reallywidehat
\usepackage{algorithm, algpseudocode}
\usepackage{setspace}
\usepackage{url} 
\usepackage{lscape}
\usepackage{booktabs} 
\usepackage{titlesec}
\usepackage{xr}
\usepackage{rotating}

\usepackage{hyperref}

\usepackage{multirow}
\usepackage{caption}

\usepackage{bbm}
\usepackage{pdflscape, array, rotating, multirow}

\usepackage{centernot}

\usepackage{lscape}

\titleformat{\subsection}
      {\normalfont\fontsize{12}{17}\bfseries}{\thesubsection}{1em}{}
      
\renewcommand{\thetable}{S\arabic{table}}
\renewcommand{\thefigure}{S\arabic{figure}}
\renewcommand{\theequation}{S\arabic{equation}}
\theoremstyle{definition}
\newtheorem{example}{Example}[section]

\def\bSig\mathbf{\Sigma}

\let\OLDthebibliography\thebibliography
\renewcommand\thebibliography[1]{
  \OLDthebibliography{#1}
  \setlength{\parskip}{0pt}
  \setlength{\itemsep}{0pt plus 3ex}
}
\newcommand\norm[1]{\left\lVert#1\right\rVert}
\makeatletter
\newcommand*{\indep}{%
  \mathbin{%
    \mathpalette{\@indep}{}%
  }%
}
\newcommand*{\nindep}{%
  \mathbin{%                   % The final symbol is a binary math operator
    \mathpalette{\@indep}{\not}% \mathpalette helps for the adaptation
  }%
}
\newcommand*{\@indep}[2]{%
  \sbox0{$#1\perp\m@th$}%        box 0 contains \perp symbol
  \sbox2{$#1=$}%                 box 2 for the height of =
  \sbox4{$#1\vcenter{}$}%        box 4 for the height of the math axis
  \rlap{\copy0}%                 first \perp
  \dimen@=\dimexpr\ht2-\ht4-.2pt\relax
  \kern\dimen@
  {#2}%
  \kern\dimen@
  \copy0 %                       second \perp
} 
\makeatother

\makeatletter
\newcommand*{\addFileDependency}[1]{% argument=file name and extension
  \typeout{(#1)}
  \@addtofilelist{#1}
  \IfFileExists{#1}{}{\typeout{No file #1.}}
}
\makeatother

\articletype{Article Type}%

\received{26 April 2016}
\revised{6 June 2016}
\accepted{6 June 2016}

\begin{document}

\title{Structured Learning in Time-dependent Cox Models
 % \protect\thanks{These two authors contributed equally to this work.}
}

\author[1]{Guanbo Wang$\dagger$*}

\author[2]{Yi Lian$\dagger$}

\author[3,4]{Archer Y. Yang*}

\author[5]{Robert W. Platt}

\author[6,7]{Rui Wang}

\author[8]{Sylvie Perreault}

\author[9]{Marc Dorais}

\author[8,10]{Mireille E. Schnitzer}

\authormark{G. Wang \textsc{et al}}

% \address[1]{\orgdiv{CAUSALab}, \orgname{Harvard T.H. Chan School of Public Health}, \orgaddress{\state{MA}, \country{U.S.A.}}}

\address[1]{\orgdiv{Department of Epidemiology}, \orgname{Harvard T.H. Chan School of Public Health}, \orgaddress{\state{MA}, \country{U.S.A.}}}

\address[2]{\orgdiv{Department of Biostatistics, Epidemiology and Informatics}, \orgname{University of Pennsylvania}, \orgaddress{\state{PA}, \country{U.S.A.}}}

\address[3]{\orgdiv{Department of Mathematics and Statistics}, \orgname{McGill University}, \orgaddress{\state{QC}, \country{Canada}}}

\address[4]{\orgdiv{Mila}, \orgname{Qu\'ebec AI Institute}, \orgaddress{\state{QC}, \country{Canada}}}

\address[5]{\orgdiv{Department of Epidemiology, Biostatistics and Occupational Health}, \orgname{McGill University}, \orgaddress{\state{QC}, \country{Canada}}}

\address[6]{\orgdiv{Department of Population Medicine}, \orgname{Harvard Pilgrim Health Care Institute and Harvard Medical School}, \orgaddress{\state{MA}, \country{U.S.A.}}}

\address[7]{\orgdiv{Department of Biostatistics}, \orgname{Harvard T. H. Chan School of Public Health}, \orgaddress{\state{MA}, \country{U.S.A.}}}

\address[8]{\orgdiv{Facult\'e de pharmacie}, \orgname{Universit\'e de Montr\'eal}, \orgaddress{\state{QC}, \country{Canada}}}

\address[9]{\orgdiv{StatSciences Inc.}\orgaddress{\state{QC}, \country{Canada}}}

\address[10]{\orgdiv{D\'epartement de m\'edecine sociale et pr\'eventive}, \orgname{Universit\'e de Montr\'eal}, \orgaddress{\state{QC}, \country{Canada}}\\
$\dagger$ co-first author}

\corres{*Guanbo Wang,  \email{gwang@hsph.harvard.edu} * Archer Y. Yang, 
\email{archer.yang@mcgill.ca}}

\presentaddress{677 Huntington Ave, Boston, MA 02115}

\abstract[Summary]{Cox models with time-dependent coefficients and covariates are widely used in survival analysis. In high-dimensional settings, sparse regularization techniques are employed for variable selection, but existing methods for time-dependent Cox models lack flexibility in enforcing specific sparsity patterns (i.e., covariate structures). We propose a flexible framework for variable selection in time-dependent Cox models, accommodating complex selection rules. Our method can adapt to arbitrary grouping structures, including interaction selection, temporal, spatial, tree, and directed acyclic graph structures. It achieves accurate estimation with low false alarm rates. We develop the \texttt{sox} package, implementing a network flow algorithm for efficiently solving models with complex covariate structures. \texttt{sox} offers a user-friendly interface for specifying grouping structures and delivers fast computation. Through examples, including a case study on identifying predictors of time to all-cause death in atrial fibrillation patients, we demonstrate the practical application of our method with specific selection rules.}


\jnlcitation{\cname{%
\author{G. Wang},
\author{Y. Lian},
\author{AY. Yang},
\author{RW. Platt}, 
\author{R. Wang}, 
\author{S. Perreault}, 
\author{M. Dorais}, and
\author{ME. Schnitzer}} (\cyear{xxxx}),
\ctitle{Structured Learning in Time-dependent Cox Models}, \cjournal{xx}, \cvol{xx}.}

\maketitle

% \footnotetext{\textbf{Abbreviations:} ANA, anti-nuclear antibodies; APC, antigen-presenting cells; IRF, interferon regulatory factor}

% Title of paper
\title{Structured Learning in Time-dependent Cox Models}
% \author{Guanbo Wang$^ {1, 2,\ast}$\email{gwang@hsph.harvard.edu}
% Yi Lian$^{3}$,
% Archer Y. Yang$^{4,5}$,
% Robert W. Platt$^{6}$,Rui Wang$^{7,8}$,
% Sylvie Perreault$^{9}$ 
% Marc Dorais$^{10}$ and
% Mireille E. Schnitzer$^{9,11}$\\
% $^{1}$CAUSALab, Harvard T.H. Chan School of Public Health, MA, U.S.A. ,
% $^{2}$Department of Epidemiology, Harvard T.H. Chan School of Public Health, MA, U.S.A. ,
% $^{3}$Department of Biostatistics, Epidemiology and Informatics, University of Pennsylvania, PA, U.S.A.,
% $^{4}$Department of Mathematics and Statistics, McGill University, QC, Canada,
% $^{5}$Mila - Qu\'ebec AI Institute, QC, Canada,
% $^{6}$Department of Epidemiology, Biostatistics and Occupational Health, McGill University, QC, Canada,
% $^{7}$Department of Population Medicine, Harvard Pilgrim Health Care Institute and Harvard Medical School, MA, USA,
% $^{8}$Department of Biostatistics, Harvard T. H. Chan School of Public Health, MA, USA,
% $^{9}$Facult\'e de pharmacie, Universit\'e de Montr\'eal, QC, Canada,
% $^{10}$StatSciences Inc., Notre-Dame-de-l'Île-Perrot, QC, Canada,
% $^{11}$D\'epartement de m\'edecine sociale et pr\'eventive,  Universit\'e de Montr\'eal, QC, Canada\\[2pt]
% {gwang@hsph.harvard.edu}}

% Running headers of paper:
\markboth%
% First field is the short list of authors
{G. Wang and others}
% Second field is the short title of the paper
{Structured Learning in Time-dependent Cox Models}

\maketitle

% \footnotetext{\textbf{Abbreviations:} ANA, anti-nuclear antibodies; APC, antigen-presenting cells; IRF, interferon regulatory factor}

% Title of paper
\title{Structured Learning in Time-dependent Cox Models}
% \author{Guanbo Wang$^ {1, 2,\ast}$\email{gwang@hsph.harvard.edu}
% Yi Lian$^{3}$,
% Archer Y. Yang$^{4,5}$,
% Robert W. Platt$^{6}$,Rui Wang$^{7,8}$,
% Sylvie Perreault$^{9}$ 
% Marc Dorais$^{10}$ and
% Mireille E. Schnitzer$^{9,11}$\\
% $^{1}$CAUSALab, Harvard T.H. Chan School of Public Health, MA, U.S.A. ,
% $^{2}$Department of Epidemiology, Harvard T.H. Chan School of Public Health, MA, U.S.A. ,
% $^{3}$Department of Biostatistics, Epidemiology and Informatics, University of Pennsylvania, PA, U.S.A.,
% $^{4}$Department of Mathematics and Statistics, McGill University, QC, Canada,
% $^{5}$Mila - Qu\'ebec AI Institute, QC, Canada,
% $^{6}$Department of Epidemiology, Biostatistics and Occupational Health, McGill University, QC, Canada,
% $^{7}$Department of Population Medicine, Harvard Pilgrim Health Care Institute and Harvard Medical School, MA, USA,
% $^{8}$Department of Biostatistics, Harvard T. H. Chan School of Public Health, MA, USA,
% $^{9}$Facult\'e de pharmacie, Universit\'e de Montr\'eal, QC, Canada,
% $^{10}$StatSciences Inc., Notre-Dame-de-l'Île-Perrot, QC, Canada,
% $^{11}$D\'epartement de m\'edecine sociale et pr\'eventive,  Universit\'e de Montr\'eal, QC, Canada\\[2pt]
% {gwang@hsph.harvard.edu}}

% Running headers of paper:
\markboth%
% First field is the short list of authors
{G. Wang and others}
% Second field is the short title of the paper
{Structured Learning in Time-dependent Cox Models}

\maketitle

% Add a footnote for the corresponding author if one has been
% identified in the author list
% \footnotetext{To whom correspondence should be addressed.}

% \begin{abstract}
% {Cox models with time-dependent coefficients and covariates are widely used in survival analysis. In high-dimensional settings, sparse regularization techniques are employed for variable selection, but existing methods for time-dependent Cox models lack flexibility in enforcing specific sparsity patterns (i.e., covariate structures). We propose a flexible framework for variable selection in time-dependent Cox models, accommodating complex selection rules. Our method can adapt to arbitrary grouping structures, including interaction selection, temporal, spatial, tree, and directed acyclic graph structures. It achieves accurate estimation with low false alarm rates. We develop the \texttt{sox} package, implementing a network flow algorithm for efficiently solving models with complex covariate structures. \texttt{sox} offers a user-friendly interface for specifying grouping structures and delivers fast computation. Through examples, including a case study on identifying predictors of time to all-cause death in atrial fibrillation patients, we demonstrate the practical application of our method with specific selection rules.}
% {Grouping structures,  high-dimensional data, network flow algorithm, time-dependent Cox models, structured sparse regularization, structured variable selection, survival analysis}
% \end{abstract}
\section{Introduction}
The Cox model \citep{reid2018analysis} is a well-established statistical model widely used for survival data analysis. Incorporating time-dependent covariates and coefficients in the Cox model offers more flexibility in representing associations between covariates and the hazard of the event of interest. Examples of time-varying covariates include medication usage \citep{holcomb2013prospective} and disease status \citep{weisz2016ranolazine}. Integrating time-varying coefficients into the Cox model is particularly relevant in cases where the relationship between covariates and the outcome of interest changes over time. %such as the Karnofsky score, and patient mortality changes over time. For instance, any Karnofsky score measurement beyond six months is considered less influential, as patients with lower scores are more likely to be lost to follow-up.

In many real-world applications of time-dependent Cox models, the number of covariates can be very large, potentially exceeding the number of observations in the data. To address the challenges of model overfitting and perform variable selection in such high-dimensional settings, sparse regularization techniques can be employed. These techniques help remove redundant covariates from the model and improve estimation/prediction accuracy. For example, LASSO and SCAD regularization methods have been extensively studied for Cox models \citep{huang2013oracle, fan2002variable}. In the context of time-dependent covariates and coefficients, some variable selection methods in the Cox model have been proposed \cite{beretta2019variable,yan2012model}. However, these methods only select variables individually and do not enforce specific sparsity patterns on the covariates.

Investigators often have prior knowledge about the structure of potential model covariates, which imposes certain restrictions on how covariates should be included in the model. For example, strong heredity states that ``if the interaction term is selected, then the main terms should also be selected'' \citep{lim2015learning}. To incorporate such information, which we refer to as ``selection rules'', a penalty can be applied to a weighted sum of the norms of group variable coefficients. Different specifications of grouping structures corresponds to different selection rules. In the context of Cox models without time-varying covariates or coefficients, various methods have been proposed to incorporate specific types of selection rules. For example, Wang et al. \cite{wang2014modified} introduced methods to incorporate strong heredity in interaction selection, while Simon et al. \cite{simon2013sparse} and Wang et al. \cite{wang2009hierarchically} developed sparse group LASSO techniques. Additionally, Dang et al. \cite{dang2021penalized} extended the latent overlapping group LASSO \citep{obozinski2011variable} to the Cox model, which requires specifying latent variables. Though the method can follow more types of selection rules, it is not scalable to high-dimensional settings with complex grouping structures due to the built-in algorithms. For instance, the method is not well studied when multi-layer groups (such as tree and graph structures) have significant overlap and the sparsity level is low \citep{villa2014proximal}. To the best of our knowledge, no method or software is available for structurally selecting variables in time-dependent Cox models.%However, there are broad categories of selection rules that cannot be followed by (latent) overlapping group LASSO, and extending previous methods to handle time-dependent covariates or coefficients is not a straightforward task.

We contribute to this field of research in several ways. First, we propose the first application of the structured sparsity-inducing penalty \citep{jenatton2011structured} to time-dependent Cox models. Our method can easily adapt to arbitrary grouping structures, allowing for the incorporation of highly complex selection rules. This flexibility enables the inclusion of various structures such as interaction selection, temporal, spatial, tree, and directed acyclic graph structures. Our estimator demonstrates low false alarm rates and high estimation accuracy.

Secondly, to reduce the computational burden caused by selecting time-dependent covariates with complex grouping structures required by our method, we develop a network flow algorithm that efficiently and effectively solves models with complex covariate structures. To ensure optimal efficiency, we implement this algorithm as an \texttt{R} package called \texttt{sox}, (stands for structured learning for time-dependent Cox) \citep{sox}, which is available on CRAN. Our software leverages established \texttt{SPAMS} packages and provides users with a user-friendly interface to specify arbitrary grouping structures. It has a \texttt{C++} core and offers fast computational speed and reliable performance.

Finally, we provide examples that illustrate how to specify grouping structures to respect complex selection rules in practical scenarios. In particular, in a case study, we apply our developed method to identify significant predictors associated with the time to death by any cause among hospitalized patients with atrial fibrillation. In this analysis, we incorporate eight selection rules and demonstrate the rationale behind specifying the corresponding grouping structure.

The rest of the paper is organized as follows. In section \ref{sec:Methods}, we introduce the proposed method. Section \ref{sec:Proximal} illustrates the algorithm for \texttt{sox}, followed by implementation details in section \ref{sec:Implementation}. We then present the results of simulation studies to compare our method to unstructured variable selection in both low and high dimensional settings in section \ref{sec:Simulation}, and present the application of \texttt{sox} in the case study in section \ref{sec:Application}. We conclude with a discussion in section \ref{sec:Discussion}.
\section{Model specification and structured penalization}\label{sec:Methods}
%\subsection{The (time-dependent) Cox model} \label{sec:TDCox}
Consider individual failure times $T_{i}$ and censoring times $C_{i}$ indexed by $i=1,...,n$. We can observe only the time to either failure or censoring, whichever comes first, i.e. $U_{i}=\min(T_{i},C_{i})$ with censoring indicator $\delta_{i}=I(T_{i}\leq C_{i})$. We also consider a possibly time-varying, $p$-vector-valued covariate process $\mathbf{X}_{i}(t)=(X_{i1}(t),\ldots,X_{ip}(t))^{\top}$. We assume noninformative censoring, i.e. upon conditioning on $\mathbf{X}_{i}(t)$, $C_{i}$ is independent of $T_{i}$. Therefore the observed data associated with $n$ individuals are $n$ triplets $\{U_{i},\delta_{i},\mathbf{X}_{i}(t)\}$
for $i=1,\ldots,n$, which we assume to be independently drawn from a common distribution. Denote by $h(t|\cdot)$ the covariate-conditional hazard function; we assume that $h_{i}(t)=h_{0}(t)\cdot\exp\{\mathbf{X}_{i}(t)^{\top}\boldsymbol{\beta}\}$,
where $h_{0}(\cdot)$ is an unspecified baseline hazard function. We begin by considering the scenario in which the coefficient vector, denoted as $\boldsymbol{\beta}\in\mathbb{R}^{p}$, is time-invariant. Subsequently, we will explore the case of time-varying coefficients.

To handle tied-events, we define an index $\ell=1,\dots, L$ for the ordered unique follow-up times in the dataset, and an ordered list $t_{1}< t_{2}< \ldots< t_{L}$  of unique time-to-event realizations. The number of tied events occurring at the $\ell$th distinct survival time is denoted by $d_{\ell}$. We can further define two index sets, $D_{\ell}$ and $R_{\ell}$, representing the subjects whose event occurred at time $t_{\ell}$ and were at risk at time $t_{j}$, respectively. 
% Being "at risk" at time $t$ means that the subject has not yet experienced the event before time $t$, and has not been censored before or at time $t$. 
Using the Breslow approximation to accommodate tied events \citep{breslow1974covariance}, the negative log partial likelihood can be approximated as
% \begin{align*}
%     L(\boldsymbol{\beta}) &\approx \prod_{j=1}^{m}\frac{\exp\left[\{\sum_{\ell \in D_{j}}\mathbf{X}_{\ell}(t_j)\} \boldsymbol{\beta} \right]}{\left[\sum_{l \in R_j}\exp\{\mathbf{X}_{\ell}(t_j) \boldsymbol{\beta}\}\right]^{d_{j}}},
% \end{align*}
% and
% =-\text{log}\{\mathcal{L}(\boldsymbol{\beta})\}& 
\begin{align}\label{eq:loglik}
    f(\boldsymbol{\beta})\approx -\sum_{\ell=1}^{L}\left( \left\{\sum_{i \in D_{\ell}}\mathbf{X}_{i}(t_\ell)^{\top}\right\} \boldsymbol{\beta} -d_{\ell} \log \left[ \sum_{i \in R_\ell}\exp\{\mathbf{X}_{i}(t_\ell)^{\top} \boldsymbol{\beta}\}\right]\right).
\end{align}
Up until now, we assume the coefficients $\boldsymbol{\beta}$ are time-invariant, but there are several ways to accommodate the case where  $\boldsymbol{\beta}$ depends on time $t$. For example, for $j=1,\dots,p$, let $\beta_j(t)$ be $\sum_{m=1}^{M}I(T_m\leqslant t < T_{m+1})\beta_{jm}$ (with specified time intervals $[T_m, T_{m+1}), m=1,\ldots,M$) or $a_j+b_j\text{log}(t)$, as functions of $t$. More flexibly, let $\beta_j(t)=\sum_{m=1}^{M}\theta_{jm}\phi_{m}(t)$, where  $\phi_{m}(\cdot)$ is a set of B-spline basis functions for approximating the function $\beta_{j}(t)$. The problem of estimating $\beta_{j}(t)$ is then transformed
to the problem of estimating $a_j, b_j, \boldsymbol{\beta}_j=(\beta_{j1},\ldots,\beta_{jm})^{\top}$, or $\boldsymbol{\theta}_{j}=(\theta_{j1},\ldots,\theta_{jM})^{\top}$. 
For the sake of simplicity, we adopt the notation of time-invariant coefficients, as given in equation \eqref{eq:loglik}, as the primary framework in this paper. But we note that Cox models with time-dependent coefficients can be viewed as a specific instance in the broader context. For a detailed explanation, please refer to Example \ref{ex:TDcoef}.

To incorporate a selection rule into selecting time-dependent covariates, one approach is to enforce the collective selection of groups of covariates. However, some previous approaches have imposed restrictions on the grouping structure \citep{kim2012analysis,simon2013sparse}, such as the prohibition of overlap between groups. In contrast, this work adopts a flexible approach by imposing no constraints on the grouping structure, allowing for the consideration of a wider range of selection rules \citep{jenatton2011structured}. Let $\mathbb{V}(t)=\{X_1(t),X_2(t),\ldots,X_p(t)\}$ denote the set containing all the covariates (for the brevity, we will use $\mathbb{V}$ throughout the paper). Suppose there are $K$ pre-defined groups of these covariates, and let us define the grouping structure as $\mathbb{G}={\mathbb{g}_{k}, k=1\dots,K}$, where $\mathbb{g}_{k}$ represents a group -- a non-empty subset of $\mathbb{V}$ -- and the union of all $\mathbb{g}_{k}$'s is equal to $\mathbb{V}$. It is worth noting that the groups can overlap, meaning that $\mathbb{g}_{j} \cap \mathbb{g}_{k}$ may not be empty for $j\neq k$. To denote a vector of the same length as $\boldsymbol{\beta}$, with non-zero entries corresponding to the covariates in $\mathbb{g}$ and zero entries elsewhere, we use $\boldsymbol{\beta}_{|\mathbb{g}}$.

To select variables according to the pre-defined groups to achieve structural selection in time-dependent Cox models, we solve the following problem
\begin{align}
\label{eq:objfun}
    \underset{\boldsymbol{\beta}}{\mathrm{min}} f(\boldsymbol{\beta})+\lambda\Omega(\boldsymbol{\beta}), \quad \Omega(\boldsymbol{\beta})=\sum_{\mathbb{g}\in \mathbb{G}}\omega_{\mathbb{g}}\norm{\boldsymbol{\beta}_{|\mathbb{g}}}_{q}.
\end{align}
Here, $f(\boldsymbol{\beta})$ is a convex differentiable function as defined in Equation \eqref{eq:loglik}. The weight $\omega_{\mathbb{g}}$ is a positive user-defined value associated with the group $\mathbb{g}$, and $\Omega(\boldsymbol{\beta})$ represents the weighted sum of sparsity-inducing $\ell_q$ norms ($q=2$ or $\infty$) applied to groups of coefficients $\boldsymbol{\beta}_{|\mathbb{g}}$, where $\mathbb{g} \in \mathbb{G}$. The choice of norm can be either the $\ell_{\infty}$ norm (which corresponds to the maximum absolute value of $\boldsymbol{\beta}_{|\mathbb{g}}$) or the $\ell_{2}$ norm. Both norms serve the purpose of  encouraging the collective selection of a group of variables. However, in this paper, we primarily focus on the piece-wise linear $\ell_{\infty}$ norm.

By employing this type of penalization, each group of variables can be excluded from the model as a group, thereby promoting sparsity. The specifications of the grouping structure $\mathbb{G}$, determining the membership of variables in each group $\mathbb{g}$, leads to different sets of variables that can be selected (i.e., the complement of the union of the groups). This allows for incorporating various a priori knowledge or structures exhibited in the real data. Importantly, we allow for the inclusion of highly overlapped groups, enabling the inclusion of a wide range of structures. 

To operationalize the structures in a mathematical and explicit manner, we initially translate them into selection rules, which represent the dependencies among variables. Subsequently, we specify the grouping structure $\mathbb{G}$ to adhere to these selection rules. This approach allows us to articulate and identify the structures to be incorporated effectively. While we do not delve into the detailed explanation in this article, interested readers can find further information in existing literature \citep{yan2017hierarchical,wang2021general}.

Various types of selection rules can be followed by the structured sparsity-inducing penalty, such as strong heredity, temporal and spatial structures, and rules that require tree or graph grouping structures. In \ref{A:examples}, we provide five detailed examples of such selection rules and their related grouping structure specifications. More examples of selection rules can be found in \citep{jenatton2011structured} and \citep{mairal2011convex}.

In the next section, we present an efficient algorithm for solving the objective function \eqref{eq:objfun} for time-dependent Cox models, allowing for the incorporation of selection rules that can be followed by the structured sparsity-inducing penalty. 

\section{Proximal gradient with network flow algorithm}
\label{sec:Proximal}
In this section, we illustrate the use of a proximal gradient algorithm \citep{boyd2004convex} with network flow to solve \eqref{eq:objfun} with a structural penalty.

Since $\Omega(\boldsymbol{\beta})$ is not differentiable on its entire support, the optimization of the penalized likelihood requires the proximal method. The proximal method \citep{moreau1962fonctions} has been successfully applied in various research areas, including signal processing \citep{combettes2011proximal} and machine learning \citep{bach2010structured}.

To address the computational challenge posed by the non-smooth component in the objective function, the proximal method updates estimates that remain close to the gradient update for the differentiable function $f(\boldsymbol{\beta})$, while also minimizing the non-differentiable penalty term \citep{beck2009fast}. This approach enjoys a linear convergence rate \citep{proximal,beck2017first}. More specifically, the updated value of $\boldsymbol{\beta}$ in each iteration of the proximal gradient algorithm, denoted as $\boldsymbol{\beta}^{+}$, is obtained by minimizing the following approximated problem. Here, the loss function $f$ is approximated by a quadratic function:
\begin{align*}
\begin{split}
%\label{prox_beta}
   \boldsymbol{\beta}^{+}
   %=& \underset{\boldsymbol{\beta}}{\mathrm{argmin}}f(\boldsymbol{\beta})+\lambda\Omega(\boldsymbol{\beta})\\
    =&\underset{\boldsymbol{\beta}}{\mathrm{argmin}} f(\tilde{\boldsymbol{\beta}})+(\boldsymbol{\beta}-\tilde{\boldsymbol{\beta}})\nabla f(\tilde{\boldsymbol{\beta}})+\frac{1}{2q}\left\|\boldsymbol{\beta}-\tilde{\boldsymbol{\beta}}\right\|_{2}^{2}+\lambda\Omega(\boldsymbol{\beta})\\
    =&\underset{\boldsymbol{\beta}}{\mathrm{argmin}} \frac{1}{2q}\left\|\boldsymbol{\beta}-\{\tilde{\boldsymbol{\beta}}-q\nabla f(\tilde{\boldsymbol{\beta}})\}\right\|_{2}^{2}+{\lambda}\Omega(\boldsymbol{\beta}),
\end{split}
\end{align*}  
where $\tilde{\boldsymbol{\beta}}$ is the value of $\boldsymbol{\beta}$ from the previous iteration, and $t$ is the step size of the update. By defining $\boldsymbol{u}=\tilde{\boldsymbol{\beta}}-q\nabla f(\tilde{\boldsymbol{\beta}})$, the above problem can be further written as 
\begin{align}\label{eq:proximal}
   \boldsymbol{\beta}^{+} = \text{prox}_{q\lambda\Omega}\left\{\tilde{\boldsymbol{\beta}}-q\nabla f(\tilde{\boldsymbol{\beta}})\right\} \coloneqq \underset{\boldsymbol{\beta}}{\mathrm{argmin}}\frac{1}{2}\norm{\boldsymbol{\beta}-\boldsymbol{u}}_{2}^{2}+q\lambda\sum_{\mathbb{g}\in \mathbb{G}}\omega_{\mathbb{g}}\norm{\boldsymbol{\beta}_{|\mathbb{g}}}_{\infty}.
\end{align}

In many cases, the proximal operator can be computed in a closed form, leading to efficient computations. However, for more complex structures such as nested groups (e.g., tree structures) or general directed acyclic graphs (e.g., Example \ref{ex:dose} and the first one in Example \ref{ex:TDcoef}), the closed-form proximal operator may not exist. In the case of a tree structure, the proximal operator can still be efficiently computed using its dual form in a blockwise coordinate ascent fashion \citep{jenatton2011proximal}. However, dealing with general directed acyclic graphs presents a greater challenge. To address this issue, Marial et al. \cite{mairal2011convex} converted the dual form of the proximal operator into a quadratic min-cost flow problem, enabling efficient computations in the context of such graphs.

Define the dual variables $\boldsymbol{\xi}_{|\mathbb{g}}$, which satisfies $\sum_{\mathbb{g}\in\mathbb{G}}\boldsymbol{\xi}_{|\mathbb{g}}=\boldsymbol{\beta}$.

The dual of problem (\ref{eq:proximal}),
\begin{align}\label{eq:dual}
    \min_{\boldsymbol{\xi}}\frac{1}{2}\norm{\boldsymbol{u}-\sum_{\mathbb{g}\in\mathbb{G}}\boldsymbol{\xi}_{|\mathbb{g}}}_{2}^{2}, \text{ s.t. } \forall\mathbb{g}\in\mathbb{G}, \norm{\boldsymbol{\xi}_{|\mathbb{g}}}_{1}\leqslant\lambda\omega_{\mathbb{g}} \text{ and } \boldsymbol{\xi}_{|\mathbb{g},j}=0 \text{ if } j\notin \mathbb{g},
\end{align} 
can be transformed into a quadratic min-cost flow problem \citep{bertsekas1998network}.  
This conversion allows us to efficiently solve the problem using a network flow algorithm based on the mini-cut theorem \citep{ford1956maximal}. The algorithm converges in a finite and polynomial number of operations, providing an effective solution to the dual problem.

The network flow algorithm, commonly used in graph models \citep{babenko2006experimental}, has found widespread application in various machine learning domains, such as in image processing \citep{mairal2014sparse}. Because of the special form of the constrain in our problem, we are able to present a more efficient version of the algorithm\citep{mairal2011convex}, referred to as Algorithm \ref{Algo_prox},
% (Table \ref{tab:Algo_prox}), 
for solving (\ref{eq:dual}). The central computation in the algorithm is the evaluation of $\sum_{\mathbb{g}\in\mathbb{G}}\boldsymbol{\xi}_{|\mathbb{g}}$, accomplished by the \texttt{computeFlow} function. The details of the algorithm's steps are presented in \ref{A:BasicSteps}.

% \begin{table}[]
%     \centering
%     \begin{tabular}{c}
% % The algorithm can be implemented by the \texttt{R} function \texttt{spams.proximalGraph}
% \noindent\begin{minipage}{\textwidth}
% \renewcommand\footnoterule{} 
\begin{algorithm}[]
\caption{Solving (\ref{eq:dual}) using quadratic min-cost flow}
\label{Algo_prox}
\textbf{Inputs:} The estimate in the $k$th step $\boldsymbol{\beta}^{k}\in \mathbb{R}^{p}$, step size $q$, $\mathbb{V}$, $\mathbb{G}$, $\omega_\mathbb{g}$, $\lambda$.
%Denote $\boldsymbol{u}=\boldsymbol{\beta}^{k}-t \nabla f(\boldsymbol{\beta}^{k})$\;\\
Set $\boldsymbol{\xi}=0$.\;\\
Compute $\sum_{\mathbb{g}\in\mathbb{G}}\boldsymbol{\xi}_{|\mathbb{g}}\leftarrow\texttt{computeFlow}(\mathbb{V},\mathbb{G})$.\\
\Return $\boldsymbol{\beta}^{k}-q \nabla f(\boldsymbol{\beta}^{k})-\sum_{\mathbb{g}\in\mathbb{G}}\boldsymbol{\xi}_{|\mathbb{g}}$\\
 \textbf{Function} $\texttt{computeFlow}(\mathbb{V},\mathbb{G})$\\
Projection: $\boldsymbol{\gamma} \leftarrow \text{argmin}_{\boldsymbol{\gamma}}\sum_{j: X_j\in\mathbb{V}}\frac{1}{2t}(\beta_j^{k}-q \nabla f(\beta_j^{k})-\gamma_j)^2 \text{ s.t. } \sum_{j: X_j\in\mathbb{V}}\gamma_j\leqslant\lambda\sum_{\mathbb{g}\in\mathbb{G}}\omega_{\mathbb{g}}$\\
Updating: $(\sum_{\mathbb{g}\in\mathbb{G}}\boldsymbol{\xi}^j_{|\mathbb{g}})_{X_j\in\mathbb{V}}\leftarrow\text{argmax}_{(\sum_{\mathbb{g}\in\mathbb{G}}\boldsymbol{\xi}^j_{|\mathbb{g}})_{X_j\in\mathbb{V}}}\sum_{X_j\in\mathbb{V}}\sum_{\mathbb{g}\in\mathbb{G}}\boldsymbol{\xi}^j_{|\mathbb{g}}
% \\ 
% \hspace{0.3cm}
\text{ s.t. }\sum_{X_j\in\mathbb{g}}\boldsymbol{\xi}^j_{|\mathbb{g}}\leqslant\lambda\omega_{\mathbb{g}} $\\
\text{Recursion:}\\
\If{$\exists X_j\in \mathbb{V} \text{ s.t. } ,  \sum_{\mathbb{g}\in\mathbb{G}}\boldsymbol{\xi}^j_{|\mathbb{g}}\neq\gamma_j$}
    {
        Denote $\mathbb{V}^{*}=\{X_j\in\mathbb{V}: \sum_{\mathbb{g}\in\mathbb{G}}\boldsymbol{\xi}^j_{|\mathbb{g}}=\gamma_j\}$, and $\mathbb{g}^{*}=\{\mathbb{g}\in\mathbb{G}: \sum_{X_j\in\mathbb{g}}\boldsymbol{\xi}^j_{|\mathbb{g}}<\lambda\omega_{\mathbb{g}}\}$\\
        $(\sum_{\mathbb{g}\in\mathbb{G}}\boldsymbol{\xi}^j_{|\mathbb{g}})_{X_j\in\mathbb{V}^{*}}\leftarrow\texttt{computeFlow}(\mathbb{V}^{*},\mathbb{G}^{*})$\\
$(\sum_{\mathbb{g}\in\mathbb{G}}\boldsymbol{\xi}^j_{|\mathbb{g}})_{X_j\in\mathbb{V}\setminus\mathbb{V}^{*}}\leftarrow\texttt{computeFlow}(\mathbb{V}\setminus\mathbb{V}^{*},\mathbb{G}\setminus\mathbb{G}^{*})$
        \;
    }
\Return $(\sum_{X_j\in\mathbb{g}}\boldsymbol{\xi}^j_{|\mathbb{g}})_{X_j\in\mathbb{V}}$
\end{algorithm}
% \end{minipage}\\\\
%     \end{tabular}
%     \caption{The algorithm for solving (\ref{eq:dual}) using quadratic min-cox flow}
%     \label{tab:Algo_prox}
% \end{table}
The algorithm is equivalent to the network flow algorithm \citep{mairal2011convex}, but with a different presentation style. Our presentation connects to the general framework for structured variable selection and is consistent with the research problem in our context.
\section{Implementation details}
\label{sec:Implementation}
We provide an efficient and user-friendly \texttt{R} package \texttt{sox}, which is available on CRAN (\url{https://cran.r-project.org/package=sox}). The statistical software is implemented in \texttt{C++} with the incorporation of programs adapted from well-established software packages including the \texttt{survival} \citep{survival-package}, \texttt{glmnet} \citep{friedman2010regularization, simon2011regularization} and \texttt{SPAMS} \citep{mairal2011convex} to ensure optimal efficiency. It provides users with a convenient interface to specify the grouping structure relevant to their specific data analysis task. In addition, the \texttt{sox} features built-in solution path and cross-validation functions with their corresponding visualization tools to facilitate model tuning and diagnostics. 

\paragraph{Details of implementing the max flow algorithm.} \hspace{0.2cm}
\label{subsec:masflow} 

We utilize the max flow algorithm, as proposed by Goldberg and Tarjan \citep{goldberg1988new}, for the efficient execution of the updating step within the \texttt{computeFlow} process. To our knowledge, it remains unmatched in terms of speed and effectiveness for solving max-flow problems. The algorithm has a worst-case complexity of $O(|V|^2|E|^{1/2})$  \citep{cherkassky1997implementing}, and is well-suited for  efficient distributed and parallel implementations. This algorithm has been effectively integrated into our package \texttt{sox}, using the \texttt{SPAMS} packages with a \texttt{C++} core. In the following section, we provide a concise introduction to this algorithm.

Consider a canonical graph where each node $v\in V$ can be a source $s_1$, and sink $s_2$, a single variable ($X_j\in\mathbb{V}$), or a set of variables ($\mathbb{g}_k\in\mathbb{G}$), that is, $V=\{s_1, s_2\}\cup\mathbb{V}\cup\mathbb{G}$. In addition, there is an arc $e\in E\subseteq V\times V$ from $s_1$, $\mathbb{g}_k$, and $X_j$ to $\mathbb{g}_k$, $X_j$, and $s_2$ respectively. From one vertex $v$ to another $w$, each arc has attributes such as non-negative functions of flow $f(v, w)$, which equals $-f(w, v))$, capacity $c(v, w)\geqslant f(v, w)$, the flow excess $h(v)=\sum_{u\in V}f(u, v)\geqslant 0, \forall v\in\{V-\{s_1\}\}$, and the residual capacity $r(v, w)=c(v, w)-f(v, w)$. See Table \ref{tab:3type} for the definitions of those functions. Therefore, the updating step in Algorithm 1 can be formulated as ``finding the maximum value of the flow while ensuring that the flow on each arc does not exceed its capacity''.

There are two basic operations in the max flow algorithm. One is \textit{push}, which pushes the excess from $v$ to $w$ by $\text{min}\{h(v), r(v, w)\}$ when $h(v)>0$ and $r(v, w)>0$. The other is \textit{relabel}, which estimates the distance from a vertex $v$ to the sink $s_2$. Define the distance as $d(v)$, where $d(s_1)=|V|$. Relabeling updates the $d(v)$ to $\text{min}\{d(w)+1|r(v, w)>0, d(v)<d(w)\}$.

The algorithm first initializes and relabels the distance, and then pushes excess from the vertex whose flow equals the capacity on each edge and can reach the sink to vertices that have a shorter estimated distance to the sink $s_2$, with the goal of getting as much excess as possible to $s_2$. When a vertex cannot reach the sink with a positive excess, the algorithm pushes such excess in the opposite direction. In each update, the value of the flow function is changed. Eventually, all vertices other than the source and sink have zero excess while each arc respects the capacity. At this point, the flow is a maximum flow, and thus, the value of the flow function can be computed as the updated value in the updating step in Algorithm 1.

% \begin{table}[H]
%     \centering
%     \begin{tabular}{llllll}
%     \hline\hline
%     \textbf{Type} & \textbf{From} & \textbf{To} & \textbf{Flow $f$} &\textbf{Capacity $c$} & \textbf{Cost $y$}\\
%     \hline
%     1 &   $s_1$ & $\mathbb{g}_k\in\mathbb{G}$ & $\sum_{X_{j}\in\mathbb{V}}\sum_{\mathbb{g}\in\mathbb{G}}\boldsymbol{\xi}^j_{|\mathbb{g}}
%     $& $\lambda\omega_{\mathbb{g}}$ & 0 \vspace{2mm} \\
%     2 & $\mathbb{g}_k$ & $X_j\in\mathbb{g}_k$ & $\boldsymbol{\xi}_{|\mathbb{g}}^{j}$ & $\infty$& 0 \vspace{2mm}\\
%     3 & $X_j\in\mathbb{V}$ & $s_2$ & $(\sum_{\mathbb{g}\in\mathbb{G}}\boldsymbol{\xi}^j_{|\mathbb{g}})_{X_j\in\mathbb{V}}$ & $\infty$& $M^{*}$ \\
%     \hline\\
%     \end{tabular}
%     \caption{Corresponding inputs of the max flow algorithm. 
%     $M^{*}=\frac{1}{2t}\{\beta_j^{k}-t \nabla f(\beta_j^{k})-(\sum_{\mathbb{g}\in\mathbb{G}}\boldsymbol{\xi}^j_{|\mathbb{g}})_{X_j\in\mathbb{V}}\}^2$}
%     \label{tab:3type}
% \end{table}
% \noindent

\paragraph{Backtracking line search.}\hspace{0.2cm} The proposed algorithm for solving the dual of the proximal operator includes the step size $q$, which enables the incorporation of backtracking line search. The algorithm, presented in Algorithm \ref{Algo_all}, allows us to solve (\ref{eq:proximal}) using proximal gradient descent with backtracking line search \citep{bertsekas1997nonlinear}. Backtracking line search is an optimization technique that helps determine the appropriate step size. It begins with a predefined step size for updating along the search direction and iteratively shrinks the step size (i.e., ``backtracks") until the decrease in the loss function corresponds reasonably to the expected decrease based on the local gradient of the loss function. This technique enhances the convergence speed of the algorithm.

% \begin{table}[]
%     \centering
%     \begin{tabular}{c}
% % The algorithm can be implemented by the \texttt{R} function \texttt{spams.proximalGraph}
% \noindent\begin{minipage}{\textwidth}
% \renewcommand\footnoterule{} 
\begin{algorithm}[]
\caption{Solving (\ref{eq:proximal}) using proximal gradient descent with backtracking line search}
\label{Algo_all}
\textbf{Inputs:} $X_i(t)$, $T_{i}$, $\delta_i$, $\mathbb{V}$, $\mathbb{G}$, $\omega_\mathbb{g}$, $\lambda$, convergence threshold $r$, shrinkage rate $\alpha <1$, step size $q$.
Set $\boldsymbol{\beta}^{0}=\mathbf{0}, k=0$.\;\\
\Repeat {$\left\|\boldsymbol{\beta}^{k}-\boldsymbol{\beta}^{k-1}\right\|_{1}<r$}{
      %$\lambda=\lambda/t$\;
      $\boldsymbol{\beta}^{+} \leftarrow \text{prox}_{q\lambda\Omega}\left(\boldsymbol{\beta}^{k}-q \nabla f(\boldsymbol{\beta}^{k})\right)$\Comment{call Algorithm \ref{Algo_prox}}\\
     \eIf{$f(\boldsymbol{\beta}^{+}) \leqslant f(\boldsymbol{\beta}^{k})+\nabla f(\boldsymbol{\beta}^{k})^{\intercal}(\boldsymbol{\beta}^{+}-\boldsymbol{\beta}^{k})+\frac{1}{2q}\left\|\boldsymbol{\beta}^{+}-\boldsymbol{\beta}^{k}\right\|_{2}^{2}$}{
     $k=k+1$;
     $\boldsymbol{\beta}^{k+1} \leftarrow \boldsymbol{\beta}^{+}$\;\\
      \textbf{exit;}
    }{
      $q \leftarrow \alpha q$\;
    } 
    }
\Return $\hat{\boldsymbol{\beta}} \leftarrow \boldsymbol{\beta}^{k+1}$
\end{algorithm}
% \end{minipage}\\\\
%     \end{tabular}
%     \caption{Solving (\ref{eq:objfun}) using proximal gradient descent with backtracking line search}
%     \label{tab:Algo_all}
% \end{table}

In our implementation, the step size shrinkage rate $\alpha$ is a parameter in the backtracking line-search that controls the rate at which the step size is reduced during each iteration of line-search until an appropriate step size is found. The proper step size should satisfy the line-search criteria, ensuring that the update with this step size leads to a sufficient decrease in the objective function. If $\alpha$ is too large, the step size might not reduce sufficiently during each iteration of the line search, which can cause the line search algorithm to take more iterations to find an appropriate step size. On the other hand, if the shrinkage factor is too small, it may lead to an over-reduction of the step size, resulting in a small update and slow convergence of the algorithm. In our implementation, we have chosen the commonly used default value of $\alpha=0.5$ as the shrinkage rate. This choice has proven to be effective for all computations in our simulations and real data analysis.

\paragraph{Cross-validation.}\hspace{0.2cm} We employ cross-validation to select the appropriate value of $\lambda$. The average cross-validated error (CV-E) is utilized for this purpose. Consider performing $L$-fold cross-validation, where we denote $\hat{\boldsymbol{\beta}}^{-l}$ as the estimate obtained from the remaining $L$-1 folds (training set). The error of the $l$-th fold (test set) is defined as $2(P-Q)/R$, where $P$ is the log partial likelihood evaluated at $\hat{\boldsymbol{\beta}}^{-l}$ using the entire dataset, $Q$ is the log partial likelihood evaluated at $\hat{\boldsymbol{\beta}}^{-l}$ using the training set, and $R$ is the number of events in the test set.

%     \item $R$ is the number of events in the test set.
% \begin{itemize}
%     \item $P$ is the log partial likelihood evaluated at $\hat{\boldsymbol{\beta}}^{-l}$ using the entire dataset;
    
%     \item $Q$ is the log partial likelihood evaluated at $\hat{\boldsymbol{\beta}}^{-l}$ using the training set;

%     \item $R$ is the number of events in the test set.
% \end{itemize} 
We opt for using the error defined above instead of the negative log partial likelihood evaluated at $\hat{\boldsymbol{\beta}}^{-l}$ using the test set because it efficiently leverages the risk set, resulting in greater stability when the number of events in each test set is small. The CV-E serves as a metric for parameter tuning. Additionally, to account for the outcome balance among randomly formed test sets, we divide the deviance $2(P-Q)$ by $R$.

\paragraph{A note on the ``One Standard Error Rule''}\hspace{0.2cm}\label{1serule.} Different values of $\lambda$ correspond to different models in the regularization framework. The selection of the appropriate $\lambda$ value is achieved through cross-validation. 

When the objective is to identify the model with the lowest prediction error, we choose the value of $\lambda$ that yields the lowest CV-E. This approach is known as the \textit{min} rule \cite{friedman2010regularization}. However, if the goal is to recover the sparsity pattern, meaning to select the set of variables that closely resembles the true model's variable set, an alternative rule called the ``one-standard-error-rule" (\textit{1se} rule) is recommended \citep{chen2021one}. The \textit{1se} rule selects the most parsimonious model whose prediction error is within one standard error of the minimum CV-E. By applying the \textit{1se} rule, we prioritize models that are more sparse while still maintaining reasonable prediction accuracy.

If the time-dependent covariates are internal covariates \citep{zhang2018time}, using the time-dependent Cox model for prediction may not be appropriate. However, when the objective is predictor identification, we recommend applying the \textit{1se} rule.

\section{Simulation}
\label{sec:Simulation}

We conduct several simulation studies. We evaluate our method in both low- and high-dimensional settings and test its performance in terms of the ability to strictly respect complex selection rules, selection consistency, estimation and prediction accuracy. We also compare our methods with the LASSO, the sparse group LASSO, and the latent overlapping group LASSO. In addition, we report the computation time, evaluate the cross-validation stability, and give insights into the influence of effect of group size, the amount of overlap, and sparsity levels on the performance of our method.
All simulations employ 10-fold cross-validation to evaluate the performance. The simulations were conducted using \texttt{R} version 4.0.5 \citep{R}. The \texttt{R} code for simulation is available at \url{https://github.com/Guanbo-W/sox_sim}.

\subsection{Categorical interaction selection under the time-dependent, low-dimensional setting}\label{subsec:sim1}
%\subsubsection{Simulation design (low-dimensional)}
% \paragraph{Covariate}
In the simulation, we generate data with three main terms, two of which are categorical variables, and two interactions between categorical variables. The three independent variables, $A(t)$, $B(t)$, and $C(t)$, are generated with values that randomly change over time in a piece-wise constant fashion. We consider 50 time points at which the values of any variable can potentially change, with each variable being held constant for a random duration between 5 and 10 time points. To simplify the notation, we use shorthand representations such as $A$ to denote $A(t)$ (similarly for $B$ and $C$). The categorical variables $A$ and $C$ are three-level variables represented by two dummy variables, denoted as $A_{1}$, $A_{2}$, $C_1$, and $C_2$, respectively. The variable $B$ is continuous.

Hence, we consider the covariates and functions of covariates as follows $\mathbf{X}=\{A_1, A_2, B, A_{1}B, A_{2}B, C_{1}, C_{2}, C_{1}B, C_{2}B\}$. The below steps outline the procedure for generating $A$ and $C$. Step 1, with replacement, sample 10 integers from \{1, 2, 3\} with equal probability to represent the three categories of the variables;
Step 2, for each sampled value, repeats the value $R_i$ times, where $R_i$ is sampled from \{5, 6, \dots, 10\} with equal probabilities. Then, concatenate these repeated values together, resulting in a single vector with a length between 50 and 100;
Step 3, take the first 50 elements as the values of the categorical variable. $B$ is generated similarly, with the only difference being that in Step 1, we generate a series of numbers from a standard normal distribution. 
% \begin{table}[]
%     \centering
%     \begin{tabular}{p{2cm}p{13cm}}
% \hline
% \textbf{Step} & \textbf{For each subject $i$}:\\\\
% Step 1 & with replacement, sample 10 integers from {1, 2, 3} with equal probability to represent the three categories of the variables.\\\\
% Step 2  &  for each sampled value, repeat the value $R_i$ times, where $R_i$ is sampled from {5, 6, \dots, 10} with equal probabilities. Then, concatenate these repeated values all together, resulting in a single vector with length between 50 and 100.\\\\
% Step 3  &  take the first 50 elements as the values of the categorical variable.\\
% \hline\\
% \end{tabular}
%     \caption{Steps for generating each time-dependent categorical variable}
%     \label{tab:Gen_A}
% \end{table}
The time-to-event outcome is generated using a permutation algorithm implemented in the \texttt{R} function \texttt{PermAlgo} \citep{PermAlgo}. The event times are dependent on the time-dependent potential predictors $\mathbf{X}$ according to the Cox model, where $\boldsymbol{\beta}^{9\times 1}$ represents the vector of log hazard ratios of the predictors. The generated data included approximately 50\% random censoring, meaning that for about half of the observations, the event time is unknown due to censoring. Among those not censored, the median event time occurred at approximately time 25.
Two scenarios were evaluated:

\begin{description}
    \item[Scenario 1:] Only $A_1$ and $A_2$ are predictive of the outcome; their coefficients are set as log(3). This represents a sparse structure.
    \item[Scenario 2:] There are five true predictors ($A_{1}$, $A_{2}$, $B$, $A_{1}B$, $A_{2}B$) that were predictive of the outcome; all of their coefficients are set as log(3). This corresponds to a less sparse structure.
\end{description}

To enforce selection rules strong heredity (selection rule 1) and ``the binary indicators representing a categorical variable are selected collectively '' (selection rule 2), we defined a grouping structure as $\big\{ \{A_{1}, A_{2}, A_{1}B, A_{2}B\}$,
$\{B, A_{1}B, A_{2}B, C_{1}B, C_{2}B\}$,
$\{A_{1}B, A_{2}B\}$,
$\{C_1, C_2, C_{1}B, C_{2}B\}$,
$\{C_{1}B, C_{2}B\}\big\}$
This grouping structure ensures that the dummy variables representing a categorical variable are selected collectively, and if an interaction term is selected, the corresponding main terms are also selected. Detailed information on how to determine the grouping structure is provided in \ref{A:GSI}.

We compare our method with the $\ell_1$-penalized Cox models with time-dependent covariates (\texttt{CoxL}, Cox LASSO) using the \texttt{glmnet} package in R.  The \texttt{glmnet} package provides an implementation of the (time-dependent) Cox model using $\ell_1$ regularization.  We use this implementation as a baseline for comparison with our method.

The performance of the two methods was evaluated using several measures, as presented in Table \ref{tab:SimMeasures}. For each simulated dataset, these measures were calculated individually and then averaged to obtain the overall performance statistics.
The weight $\omega_\mathbb{g}$ for each group in the penalization term is set to one. We use a convergence criterion of $10^{-5}$, which is based on the sum of the absolute differences between the estimates from the two steps. In the process of model selection, we compare both the \textit{min} and \textit{1se} rules for choosing the tuning parameter $\lambda$. The results are given in Table \ref{tab:SimResults}.

Our \texttt{sox} method always followed both of the rules as designed. However, Cox LASSO with the \textit{min} rule fails to respect the categorical selection rule in 20\%-40\% of the cases, and it violates the strong heredity rule in 80\% of the cases in certain settings, even with increased sample sizes. It is important to note that in scenario 2, Cox LASSO had a higher probability of breaking the rules.
When applying the \textit{min} rule, both methods achieved a perfect missing rate. Additionally, it is observed that \texttt{sox} converged faster than Cox LASSO. However, when applying the \textit{1se} rule, the missing rate of \texttt{sox} converged faster compared to Cox LASSO. This confirms that \texttt{sox} had a higher chance of successfully selecting the variables that should be selected, even with a relatively small sample size.

For both methods, the false alarm rate was much worse when using the \textit{min} rule, indicating that the methods select a significant number of noise variables. Therefore, in a low-dimensional setting, if the objective is to recover the sparsity pattern, it is advisable to avoid applying the \textit{min} rule. However, it is worth noting that the false alarm rate of \texttt{sox} with the \textit{1se} rule converged to 0 relatively quickly with increasing sample sizes. 
The MSE of \texttt{sox} was significantly smaller and converged faster compared to Cox LASSO under either rule, indicating that \texttt{sox} achieved better estimation performance. The cross-validated errors and prediction accuracy of both methods are similar across all the settings.

Overall, the simulation results verify that \texttt{sox} can more effectively recover the sparsity pattern and provide better estimation when incorporating the correct selection rules.

\subsection{Interaction selection under time-dependent, high-dimensional setting}

We follow the design outlined in She et al. \citep{she2018group} to perform interaction selection. In this model, our goal is to identify significant two-way interactions from all the potential ones while adhering to the strong heredity selection rule. We generate time-dependent predictors denoted as $X=(X^{\text{main}},X^{\text{inter}})$. Within $X^{\text{main}}=(X_{1},\ldots,X_{p'})$, each main term is generated following a standard normal distribution, similar to the low-dimensional case (e.g., $B(t)$). We introduce four time-varying points, with each variable held constant for two or three time-points. Subsequently, we create $X^{\text{inter}}=(X_{1}X_{2},\ldots,X_{1}X_{p'},X_{2}X_{3},\ldots,X_{p'-1}X_{p'})$. The true coefficient vector is denoted as
\[\boldsymbol{\beta}=(\beta_{1},\ldots,\beta_{p'},\beta_{1,2},\ldots,\beta_{1,p'},\\ \beta_{2,3},\ldots,\beta_{p'-1,p'})^{\top}\]
which has a length of $p'+\binom{p'}{2}$, matching the dimension of the combined predictors in $X$.

We generate time-to-event outcomes using the \texttt{R} package \texttt{PermAlgo}
\citep{PermAlgo}. For all simulations with different combinations of $n$ and
$p$, we set the nine coefficients of the main terms $\beta_{j}, \forall j=1,\ldots,9$ as 0.4 and the nine coefficients of interaction terms $\beta_{j,j'}=0.3$ for
$(j,j')=(1,2)$, $(1,3)$, $(1,7)$, $(1,8)$, $(1,9)$, $(4,5)$, $(4,6)$, $(7,8)$ and $(7,9)$. All other coefficients are set to zero.

To enforce the strong heredity selection rule, we define the grouping structure as follows:
$$
\mathbb{g}_{1} =\{X_{1},X_{1}X_{2},\ldots,X_{1}X_{p'}\},..., 
\mathbb{g}_{p'}  =\{X_{p'},X_{1}X_{p'},\ldots,X_{p'-1}X_{p'}\}, \quad
\mathbb{g}_{p'+1}  =\{X_{1}X_{2}\},..., \mathbb{g}_{p'+\binom{p'}{2}} =\{X_{p'-1}X_{p'}\}.    
$$
We consider six different combinations of $(n,p)$ with $n=(400,800)$ and $p=p'+\binom{p'}{2}=(210,465,820)$, where $p'=(20,30,40)$ represents the number of main terms. Since \texttt{SGL} and \texttt{grpCox} do not support the time-dependent model, we only compare the performance of \texttt{sox} with the LASSO regularized Cox model (both using the \textit{min} rule). The optimal value of $\lambda$ was selected from a sequence of candidate values through 10-fold cross-validation. We also applied adaptive penalty weights to obtain debiased estimates (\texttt{db}). Specifically, the adaptive regularization weights were calculated as the inverse of the original \texttt{sox} or the LASSO Cox estimates $\omega_{\mathbb{g}}=1/\max(|\hat{\boldsymbol{\beta}}_{\mathbb{g}}|, 10^{-16})$.

% Estimation accuracy was evaluated using the Mean Squared Error (MSE), calculated as $\|\boldsymbol{\beta}-\hat{\boldsymbol{\beta}}\|_{2}^{2}/(p'+\binom{p'}{2})$. Additionally, we reported the C-index and CV-E. 

We assess selection consistency using the same metrics as in Section 5.1. The results of these evaluations are presented in Table \ref{tab:Simulation-high-d}. In summary, \texttt{sox} outperforms CoxL in several aspects. Firstly, \texttt{sox} strictly adheres to the strong heredity selection rule, which is not achieved by CoxL. Secondly, \texttt{sox} exhibits lower missing rates than CoxL when $n=400$ and comparable missing rates when $n=800$. Debiased \texttt{sox} has the lowest false alarm rates in most cases. This can be attributed to the enforcement of the selection rule, where the elimination of an interaction term from the model is triggered not only by the term itself but also by the absence of either of its main terms. Furthermore, \texttt{sox} demonstrates higher estimation accuracy, as evidenced by the lower mean-squared errors.

When incorporating the strong heredity selection rule, the algorithm forces the selection of the two main terms when an interaction is selected, which would increase the false alarm rate if the selected interaction term is a noisy variable. Therefore, the false alarm rate of \texttt{sox} can be slightly higher than the methods without incorporating the selection rule. However, by employing our method, we can achieve a lower missing rate and more importantly, an interpretable prediction model.

Similar conclusions can be made when the \textit{1se} rule is applied. The results are given in \ref{app:add_5.2}.

\subsection{Comparison with existing sparse group lasso methods}

In this section, we compare \texttt{sox} with two existing packages, \texttt{SGL} \citep{sgl-package} and \texttt{grpCox} \citep{grpCox-package}, both of which implement the sparse group lasso for the time-fixed Cox model. Specifically, \texttt{grpCox} achieves within-group sparsity by utilizing a latent group LASSO approach \citep{jacob2009group,obozinski2011group}.

Following Simon et al. \citep{simon2013sparse}, we simulate a covariate matrix $\mathbf{X}$ with dimensions $n=100$ and $p=200$. The columns of $\mathbf{X}$ are independently generated from a standard Gaussian distribution. The variables $X_{1},\ldots,X_{200}$ are divided into 10 groups, each containing 20 variables.

We consider three different cases of true coefficients 
\begin{eqnarray*}
\text{Case 1: }\qquad\boldsymbol{\beta}&=&(\boldsymbol{\beta}_{s}^{\top},0,\ldots,0)^{\top},\\
\text{Case 2: }\qquad\boldsymbol{\beta}&=&(\boldsymbol{\beta}_{s}^{\top},\boldsymbol{\beta}_{s}^{\top},0,\ldots,0)^{\top},\\
\text{Case 3: }\qquad\boldsymbol{\beta}&=&(\boldsymbol{\beta}_{s}^{\top},\boldsymbol{\beta}_{s}^{\top},\boldsymbol{\beta}_{s}^{\top},0,\ldots,0)^{\top},
\end{eqnarray*}
where
\[\boldsymbol{\beta}_{s}=(0.1, 0.2, 0.3, 0.4, 0.5,\underbrace{0,\ldots,0}_{15})^{\top}\in \mathbb{R}^{20}\]
We generate the
time-to-event outcome using the \texttt{R} package \texttt{coxed}
\citep{kropko2020coxed}, which employs uses duration-based simulation methods.
The generated event times  depends on fixed predictors according
to the proportional hazards model $h(t|\mathbf{X})=h_0(t)\exp(\boldsymbol{\beta}^\top\mathbf{X})$.

To fit the sparse group LASSO-regularized Cox model in \texttt{sox}, we specify
the following grouping structures:
\[
\begin{array}{cccc}
\mathbb{g}_{1}=\{X_{1}\}, & \cdots & \mathbb{g}_{20}=\{X_{20}\}, & \mathbb{g}_{201}=\{X_{1},\ldots,X_{20}\},\\
\mathbb{g}_{21}=\{X_{21}\}, & \cdots & \mathbb{g}_{40}=\{X_{40}\}, & \mathbb{g}_{202}=\{X_{21},\ldots,X_{40}\},\\
\vdots &  & \vdots & \vdots\\
\mathbb{g}_{181}=\{X_{181}\}, & \cdots & \mathbb{g}_{200}=\{X_{200}\}, & \mathbb{g}_{210}=\{X_{181},\ldots,X_{200}\},
\end{array}
\]
where each group $\mathbb{g}_{1},\ldots,\mathbb{g}_{200}$ contains only a single covariate
with the corresponding index, and each group $\mathbb{g}_{201},\ldots,\mathbb{g}_{210}$ contains 20 covariates. The groups adhere to a nested structure, e.g. $\mathbb{g}_{1},\ldots,\mathbb{g}_{200}$ are nested within group $\mathbb{g}_{201}$ and $\mathbb{g}_{21},\ldots,\mathbb{g}_{400}$ within $\mathbb{g}_{202}$, and so on.
The grouping structure is built for \texttt{grpCox} and \texttt{SGL}. For \texttt{SGL}, the mixing parameter (of the $\ell_{1}$
and $\ell_{2}$ component) in \texttt{SGL} is set to 0.5. For a fair comparison, we ensure that the amount of regularization applied by \texttt{sox} is effectively the same as in \texttt{SGL}. Specifically, for groups $\mathbb{g}_{1},\ldots,\mathbb{g}_{200}$, we set the regularization weight to 0.5, and for groups $\mathbb{g}_{201},\ldots,\mathbb{g}_{210}$, the regularization weight is set to $\sqrt{20} \times 0.5$, where 20 is the group size. In contrast, \texttt{grpCox} does not support custom regularization weights, so we use the package defaults.

We perform 10-fold cross-validation to select the optimal regularization coefficient $\lambda$ and evaluate the performance of the three methods. To ensure a fair comparison, we ensure that all methods use the same $\lambda$ sequence and the same train-validation split. In the case of \texttt{grpCox}, we employ the default settings for model tuning.

We summarize the performance statistics in Table \ref{tab:Simulation-nested}. In summary, \texttt{sox} and \texttt{SGL} perform similarly in fitting SGL-regularized time-fixed Cox models. In fact, the two sets of estimates are very close, as shown in the sample solution path in \ref{app:samplepath}. On the other hand, \texttt{grpCox} exhibits a considerably more conservative approach to variable selection, as evidenced by MR and FAR. Additionally, the mean squared errors (MSE) of  \texttt{grpCox} estimates are also higher.

Furthermore, we perform additional simulations that compare \texttt{sox} and \texttt{SGL} under some simulation settings where \texttt{SGL} failed to adhere to the selection rule. The details are provided in \ref{app:5.3add}.
\subsection{Additional simulations}
Here, we present the additional simulations.
\paragraph{Timing} 
We test the computational speed of \texttt{sox}. We adopt the simulation setting from Section 5.2. We report the computation time for solving 10-fold cross-validation on the same $\lambda$ sequence of length 30. We find that for a complex case where the sample size is $n=800$ and the number of variables is $p=820$, a comprehensive analysis can be completed in approximately 10 minutes. In contrast, for a simpler scenario with a sample size of $n=400$ and $p=210$ variables, the analysis requires less than 30 seconds to finish. More details are given in \ref{app:time}.
\paragraph{Stability of Cross-validation}
To evaluate the stability of the CV of \texttt{sox}, we conduct additional simulations using the simulation setting from Section 5.2. We simulate a single set of data $(n=400, p=210)$ and performed 10-fold CV twenty times. To demonstrate the stability of the CV of \texttt{sox}, we report the average CV error and MSE (resulting from the final model chosen by each repeated CV procedure), and their corresponding standard errors. We compare these results to those generated by \texttt{glmnet}. The mean CV error (with standard error) was 1.82 (0.03) for \texttt{sox} and 1.87 (0.03) for \texttt{glmnet}. The mean MSE$\times 10^{-3}$ (with standard error) was 8 (0.4) for \texttt{sox} and 9 (0.4) for \texttt{glmnet}. The results show that the CV procedure in \texttt{sox} is stable.
\paragraph{Comparison of different group sizes, the amount of overlap, and the sparsity levels} We conduct an additional simulation to compare the effect of different group sizes, the amount of overlap, and the sparsity levels on the performance of our method. We demonstrate that depending on the true data-generating mechanism, these factors may influence the performance of \texttt{sox}. The details are given in \ref{app:add_groupsize}.

\section{Case study: time-depended predictor identification for time to death by any cause  among patients hospitalized for atrial fibrillation}\label{sec:Application}
\paragraph{Background}\hspace{0.25cm} Atrial fibrillation (AF) is a medical condition characterized by an irregular heartbeat. Patients with AF are at a higher risk of experiencing cardiovascular complications and mortality \citep{lip2007management}. Therefore, it is important to identify predictors that contribute to these outcomes and develop a predictive model that can help understand the disease and support clinical decision-making.

In treating AF, most patients are prescribed oral anticoagulants (OAC) for long-term management. OAC includes medications such as warfarin and direct OAC (DOAC), the latter including Dabigatran, Apixaban, and Rivaroxaban, each of which can be taken as high or low dose. However, the use of OAC in AF patients is often intermittent due to contraindications, adverse effects, and the need for surgeries \citep{angiolillo2021antithrombotic}. Furthermore, the use of other medications and changes in disease conditions may vary over time among AF patients. Taking into account such time-varying information can be beneficial in identifying predictors of clinical outcomes in AF patients \citep{claxton2018new}.

\paragraph{Data}\hspace{0.25cm}In this study, we utilize the \texttt{sox} method to identify significant baseline and time-varying predictors associated with the time to all-cause death among patients hospitalized for AF who initiated OAC between 2010 and 2017 in the province of Quebec, Canada. The data used in this study were obtained from a larger dataset \citep{perreault2020oral} and represent 36,381 patients who were followed up for a period of 365 days. To ensure the validity of the analysis, we applied specific inclusion and exclusion criteria to the dataset, which are detailed in  \ref{A:exclusion}. Among the included patients, a total of 4,384 individuals experienced the event of interest (i.e., death by any cause) during the follow-up period, accounting for approximately 12.05\% of the population.

In this study, we selected a total of 24 candidate predictors based on previous research findings \citep{fauchier2015cause,fauchier2016causes,lee2018mortality,harb2021cha2ds2} and data availability. Out of these predictors, 7 are baseline (time-invariant) covariates. The baseline covariates include age, sex, medical scores, comorbidities (eight conditions), OAC use information, concomitant medication use (four drugs), and the interactions between each concomitant medication and DOAC. Additionally, we define five time-dependent indicator covariates to capture OAC use: DOAC, Apixaban, Dabigatran, OAC, and High-dose-DOAC.

The definitions and summary statistics for the covariates are provided in \ref{A:CovariateDefinition} and \ref{A:Tableone}, respectively. It is worth noting that the time-dependent covariates, especially those related to OAC use, exhibited significant changes over time. This highlights the importance of considering the time-varying nature of these covariates, as neglecting their changes may introduce substantial bias in covariates selection and coefficient estimation.

\paragraph{Analysis}\hspace{0.25cm}To assess the associations between the covariates and the time to the event of interest, we conduct both univariate and multivariate time-dependent Cox models. The crude hazard ratios from the univariate models and the adjusted hazard ratios from the multivariate models, along with their corresponding 95\% confidence intervals, are presented in  \ref{A:Cox}.

For this data analysis, we establish selection rules (including strong heredity and the rules for coefficients interpretability) to ensure the interpretability of the model. The specific selection rules are outlined in Table \ref{tab:SelectionRule}.
% \begin{table}[]
%     \centering
%     \begin{tabular}{p{1cm}p{13cm}}
%     \hline
%     \# & \textbf{Selection rule} \\
%     \hline\hline
%     1  & If Apixaban is selected, then select DOAC \\
%     2  & If Dabigatran is selected, then select DOAC \\
%     3  & If DOAC is selected, then select OAC \\
%     4  & If High-dose-DOAC is selected, then select DOAC \\
%     5...8 & If the interaction of DOAC and a concomitant medication is selected, then both DOAC and the concomitant medication are selected.\\
%     \hline\\
%     \end{tabular}
%     \caption{Selection rules}
%     \label{tab:SelectionRule}
% \end{table}
It aim to capture the following associations: 1) The use of OAC versus non-use;
2) The use of DOAC compared to warfarin;
3) The use of Apixaban compared to Rivaroxaban;
4) The use of Dabigatran compared to Rivaroxaban;
5) The use of high-dose-DOAC compared to low-dose-DOAC;
6) The simultaneous use of concomitant medication and DOAC.

The selection rules ensure that the aforementioned comparisons are estimable and correspond to the coefficients of the selected variables if selected. Further explanation and rationale for these selection rules can be found in  \ref{A:ApplicationSelectionRuleExplain}. To respect these selection rules simultaneously, we identified the appropriate graph-structured grouping structure, which is provided in \ref{A:GroupingStructure}, following the approach described in \citep{wang2021general,wang2022structured}.

\paragraph{Results}\hspace{0.25cm}First, we applied both the \texttt{sox} method and Cox LASSO to select variables from the data. We then used the unpenalized time-dependent Cox model to estimate the hazard ratios for the selected covariates. Additionally, we report the estimates with 95\% confidence intervals when all the covariates are included in the model. The results are given in Table \ref{tab:ApplicationResults}.
Our proposed method, \texttt{sox}, identified 15 predictors that are associated with the outcome. Comparing Cox LASSO with \texttt{sox}, we observe the following differences: 1) DOAC was not selected by Cox LASSO, which affects the interpretation of the coefficient of Apixaban; 2) the interaction of DOAC and Statin was selected without Statin in Cox LASSO, violating the strong heredity assumption; 3) predictors such as High-dose-DOAC and Beta-Blockers were not selected by Cox LASSO. Additionally, our method achieves a slightly higher concordance index compared to the unstructured penalization, though slightly lower than the full model. These results demonstrate the advantages of utilizing \texttt{sox} over Cox LASSO, highlighting the benefits of incorporating prior knowledge about the underlying structure of the data. To further illustrate the results, we visualize the impact of time-dependent covariates on the survival probability. See more details in \ref{A:ApplicationVisual}. 

\section{Discussion}
\label{sec:Discussion}
In both low and high-dimension settings, incorporating a priori knowledge of covariate structures can achieve robust and interpretable models. In survival models, especially when the covariates and coefficients are time-dependent, no clear guidance and methods are available for the incorporation of such prior information. In this paper, we introduced \texttt{sox}, a novel structured sparsity-inducing penalty for the time-dependent Cox model. The method can accommodate a wide range of a priori knowledge about the data structure in the form of restrictions on covariate inclusion. We empirically showed that incorporating correct selection rules can improve model selection performance and accuracy of estimation. The developed algorithm converges fast and is able to handle high-dimensional data, which can be implemented in the developed \texttt{R} package. Through examples, simulations, and the case study, we also explored how to set appropriate selection rules and the corresponding grouping structures for the developed method in practice, which provided users with guidance on the implementation of the methods in their own application.

We would also like to emphasize that our work primarily focuses on scenarios where practitioners have prior knowledge of the definition of each variable and its relationship with others, aiming to produce an interpretable prediction model. For instance, variables may be defined as interactions of other variables to ensure the interpretability of the resulting coefficients, necessitating the use of strong heredity (as demonstrated in various examples in Web Appendix A and the case study).

In practical situations where the definition of a variable or its relationship with other variables is unclear, we recommend not including the variable in a selection rule or performing sensitivity analysis. In such cases, it is prudent to use more conservative methods like the LASSO. Incorporating uncertain or incorrect selection rules can potentially degrade the performance of the resulting model.

There are several avenues for future work. For instance, one could relax the assumption that the hazard is dependent on the current value of the covariates, assume event times follow a parametric distribution by using accelerated failure time models \citep{kalbfleisch2011statistical}, generalize to different penalty types (for example, minimax concave penalty, \cite{zhang2010nearly}, or smoothly clipped absolute deviation \cite{fan2001variable}), and investigate the impact of applying different weighting schemes.  In addition, it would be beneficial to integrate structured variable selection into causal inference, especially when the confounders or effect modifiers are high-dimensional \citep{siddique2019causal,wang2020estimating,liu2022modeling,wang2023evaluating,wang2023review,bouchard2022predictive}.

\section*{Software}
The \texttt{R} package developed for the proposed method is available at \url{https://cran.r-project.org/package=sox}. The \texttt{R} code for simulation is available at \url{https://github.com/Guanbo-W/sox_sim}.

\section*{Supplementary Material}
The reader is referred to the on-line Supplementary Materials for technical appendices.

\section*{Acknowledgments}
We would like to thank the R\'egie de l'Assurance Maladie du Qu\'ebec (RAMQ) and Ministry of Health and Social Services (MSSS) (Quebec, Canada) for providing assistance in handling the data and the Commission d’accès à l’information for authorizing the study. G. Wang is supported by the Fonds de Recherche du Québec Santé (FRQS-272161).
AY. Yang is supported by the Natural Sciences and Engineering Research Council of Canada (NSERC) Discovery Grant RGPIN-2016-05174 and Fonds de Recherche du Qu\'{e}bec Nature et Technologies Team Grant (FRQNT-327788). ME. Schnitzer is supported by a Canadian Institutes of Health Research Canada (CIHR) Research Chair tier 2 and an NSERC Discovery Grant. RW. Platt is supported by CIHR Foundation Grant (FDN-143297). 
The study was supported by the Heart and Stroke Foundation of Canada (G-17-0018326) and the Réseau Québécois de Recherche sur les Médicaments (RQRM). Please refer to the following \url{https://www.heartandstroke.ca/} and \url{http://www.frqs.gouv.qc.ca/en/}.
{\it Conflict of Interest}: None declared.

\bibliographystyle{NJDnatbib}
\bibliography{refs}

\begin{table}[!p]
    \centering
    \begin{tabular}{llllll}
    \hline\hline
    \textbf{Type} & \textbf{From} & \textbf{To} & \textbf{Flow $f$} &\textbf{Capacity $c$} \\
    \hline
    1 &   $s_1$ & $\mathbb{g}_k\in\mathbb{G}$ & $\sum_{X_{j}\in\mathbb{V}}\sum_{\mathbb{g}\in\mathbb{G}}\boldsymbol{\xi}^j_{|\mathbb{g}}
    $& $\lambda\omega_{\mathbb{g}}$  \vspace{2mm} \\
    2 & $\mathbb{g}_k$ & $X_j\in\mathbb{g}_k$ & $\boldsymbol{\xi}_{|\mathbb{g}}^{j}$ & $\infty$ \vspace{2mm}\\
    3 & $X_j\in\mathbb{V}$ & $s_2$ & $(\sum_{\mathbb{g}\in\mathbb{G}}\boldsymbol{\xi}^j_{|\mathbb{g}})_{X_j\in\mathbb{V}}$ & $\infty$ \\
    \hline\\
    \end{tabular}
    \caption{Corresponding inputs of the max flow algorithm.}
    \label{tab:3type}
\end{table}

\begin{table}[!p]
    \centering
    \begin{tabular}{p{1cm}p{13cm}}
    \hline\hline
    \textbf{\#} & \textbf{Measurements}\\
    \hline
        % (1) & Joint detection rate (JDR): an indicator of whether the selected model included all true predictors (but also possibly selected some other noise variables).\\\\
	(1)  & Missing rate (MR): the percentage of variables not selected among the true predictors.\\
	(2)  & False alarm rate (FAR) : the percentage of selected variables among the noise variables.\\
	(3)  & Rule 1 Satisfaction (R1S): whether the selected model satisfied strong heredity.\\
	(4)  & Rule 2 Satisfaction (R2S): whether the resulting selected model satisfied selection rule 2, that dummy indicators of the same variable are always selected together.\\
	(5)  & C index (RCI): the C index of the selected model.\\
	(6)  & Mean-squared error (MSE): the mean of the squared differences between each coefficient in the data generating mechanism and its estimate, i.e., the $\ell_2$-norm of the difference between the coefficient vector and its estimate.\\
	(7)  & Cross-validated error (CV-E): the cross-validated error defined in Section \ref{sec:Implementation}.\\
 \hline\\
    \end{tabular}
    \caption{Measures of comparison in the the simulation studies}
    \label{tab:SimMeasures}
\end{table}

\begin{table}[!p]
    \centering
    \begin{tabular}{l|llll|llll}
        \hline\hline
    \textbf{Scenario} & \multicolumn{4}{c}{\textbf{1 (A$\neq$0)}} & \multicolumn{4}{c}{\textbf{2 (A, B, AB$\neq$0)}}  \\\hline
    \textbf{Method} &
    \scalebox{0.85}{\texttt{sox.1se}} &
    \scalebox{0.85}{\texttt{sox.min}} & 
    \scalebox{0.85}{\texttt{CoxL.1se}} &
    \scalebox{0.85}{\texttt{CoxL.min\hspace{1pc}} }
    &\scalebox{0.85}{\texttt{sox.1se} }
    &\scalebox{0.85}{\texttt{sox.min} }
    & \scalebox{0.85}{\texttt{CoxL.1se} }
    &\scalebox{0.85}{\texttt{CoxL.min}}\\
    \hline
    & \multicolumn{8}{c}{$N=100$}\\\hline
    % \textbf{JDR} & 0.18 & 0.98 & 0.00 & 0.66 & 0.90 & 1.00 & 0.02 & 0.44 \\
    \textbf{MR} & 0.82 & 0.02 & 0.85 & 0.26 & 0.08 & 0.00 & 0.41 & 0.16 \\
    \textbf{FAR} & 0.02 & 0.65 & 0.03 & 0.40 & 0.04 & 0.53 & 0.06 & 0.49 \\
    \textbf{R1S} & 1.00 & 1.00 & 0.88 & 0.70 & 1.00 & 1.00 & 0.84 & 0.72 \\
    \textbf{R2S} & 1.00 & 1.00 & 0.75 & 0.38 & 1.00 & 1.00 & 0.39 & 0.35 \\
    \textbf{RCI} & 0.53 & 0.67 & 0.54 & 0.64 & 0.91 & 0.92 & 0.91 & 0.92 \\
    \textbf{MSE} & 0.24 & 0.06 & 0.27 & 0.14 & 0.27 & 0.10 & 0.41 & 0.30 \\
    \textbf{CV-E} & 6.87 & 6.74 & 6.68 & 6.59 & 4.70 & 4.36 & 4.74 & 4.33 \\\hline
   & \multicolumn{8}{c}{$N=500$}\\\hline
    % \textbf{JDR} & 0.88 & 1.00 & 0.50 & 1.00 & 1.00 & 1.00 & 0.08 & 0.98 \\
    \textbf{MR} & 0.12 & 0.00 & 0.32 & 0.00 & 0.00 & 0.00 & 0.36 & 0.04 \\
    \textbf{FAR} & 0.02 & 0.70 & 0.01 & 0.54 & 0.00 & 0.15 & 0.05 & 0.59 \\
    \textbf{R1S} & 1.00 & 1.00 & 0.90 & 0.57 & 1.00 & 1.00 & 0.94 & 0.78 \\
    \textbf{R2S} & 1.00 & 1.00 & 0.79 & 0.22 & 1.00 & 1.00 & 0.45 & 0.62 \\
    \textbf{RCI} & 0.60 & 0.63 & 0.58 & 0.63 & 0.91 & 0.91 & 0.91 & 0.91 \\
    \textbf{MSE} & 0.14 & 0.02 & 0.22 & 0.03 & 0.14 & 0.04 & 0.31 & 0.07 \\
    \textbf{CV-E} & 6.79 & 6.70 & 6.81 & 6.68 & 4.38 & 4.26 & 4.40 & 4.26 \\\hline
    & \multicolumn{8}{c}{$N=1000$}\\\hline
    % \textbf{JDR} & 1.00 & 1.00 & 0.94 & 1.00 & 1.00 & 1.00 & 0.20 & 0.98 \\
    \textbf{MR} & 0.00 & 0.00 & 0.04 & 0.00 & 0.00 & 0.00 & 0.20 & 0.02 \\
    \textbf{FAR} & 0.02 & 0.73 & 0.02 & 0.58 & 0.00 & 0.08 & 0.04 & 0.66 \\
    \textbf{R1S} & 1.00 & 1.00 & 0.97 & 0.70 & 1.00 & 1.00 & 0.90 & 0.82 \\
    \textbf{R2S} & 1.00 & 1.00 & 0.94 & 0.31 & 1.00 & 1.00 & 0.53 & 0.65 \\
    \textbf{RCI} & 0.61 & 0.62 & 0.60 & 0.62 & 0.91 & 0.91 & 0.91 & 0.91 \\
    \textbf{MSE} & 0.10 & 0.01 & 0.15 & 0.02 & 0.12 & 0.02 & 0.29 & 0.08 \\
    \textbf{CV-E} &6.76 & 6.68 & 6.76& 6.67 & 4.33 & 4.25& 4.36 & 4.24 \\\hline
   & \multicolumn{8}{c}{$N=2000$}\\\hline
    % \textbf{JDR} & 1.00 & 1.00 & 1.00 & 1.00 & 1.00 & 1.00 & 0.74. & 1.00 \\
    \textbf{MR} & 0.00 & 0.00 & 0.00 & 0.00 & 0.00 & 0.00 & 0.08 & 0.00 \\
    \textbf{FAR} & 0.00 & 0.06 & 0.00 & 0.54 & 0.00 & 0.00 & 0.04 & 0.57 \\
    \textbf{R1S} & 1.00 & 1.00 & 1.00 & 0.60 & 1.00 & 1.00 & 0.93 & 0.76 \\
    \textbf{R2S} & 1.00 & 1.00 & 0.99 & 0.20 & 1.00 & 1.00 & 0.79 & 0.57 \\
    \textbf{RCI} & 0.61 & 0.61 & 0.61 & 0.61 & 0.91 & 0.91 & 0.91 & 0.91 \\
    \textbf{MSE} & 0.06 & 0.00 & 0.10 & 0.01 & 0.12 & 0.05 & 0.23 & 0.01 \\
    \textbf{CV-E} & 6.71 & 6.66 & 6.72 & 6.66& 4.30 & 4.23 & 4.31 & 4.23 \\
    \hline
    \end{tabular}
    \caption{Simulation results of Section 5.1. Cox LASSO: unstructured $\ell_1$ penalty (\texttt{glmnet} with cox); \texttt{sox}: our method; \texttt{1se}: applying the 1se rule; \texttt{min}: applying the min rule. MR: Missing rate, FAR: False alarm rate, R1S: Rule 1 Satisfaction, R2S: Rule 2 Satisfaction, RCI: the C index of the model with the selected variables, MSE: Mean-squared error, CV-E: cross-validated error}
    \label{tab:SimResults}
\end{table}

\begin{table}[!p]
\centering
\begin{tabular}{cccccccccc}
 &  &  &  &  &  &  &  &  & \tabularnewline
\hline 
\hline 
Method & \texttt{sox} & \texttt{sox.db} & \texttt{CoxL} & \texttt{CoxL.db} &  & \texttt{sox} & \texttt{sox.db} & \texttt{CoxL} & \texttt{CoxL.db}\tabularnewline
\hline 
$p=210$ & \multicolumn{4}{c}{$n=400$} &  & \multicolumn{4}{c}{$n=800$}\tabularnewline
\hline 
% JDR & 0.50 & 0.35 & 0.40 & 0.25 &  & 0.75 & 0.45 & 0.80 & 0.80\tabularnewline
MR & 0.04 & 0.06 & 0.05 & 0.07 &  & 0.02 & 0.04 & 0.01 & 0.01\tabularnewline
FAR & 0.27 & 0.11 & 0.21 & 0.18 &  & 0.14 & 0.03 & 0.20 & 0.19\tabularnewline
R1S & 1.00 & 1.00 & 0.87 & 0.89 &  & 1.00 & 1.00 & 0.88 & 0.89\tabularnewline
RCI & 0.88 & 0.86 & 0.89 & 0.88 &  & 0.85 & 0.83 & 0.85 & 0.85\tabularnewline
MSE{*} & 4.97 & 3.91 & 5.97 & 4.86 &  & 4.37 & 3.88 & 5.21 & 4.11\tabularnewline
CV-E & 1.74 & 1.70 & 1.76 & 1.60 &  & 1.68 & 1.67 & 1.69 & 1.61\tabularnewline
\hline 
$p=465$ & \multicolumn{4}{c}{$n=400$} &  & \multicolumn{4}{c}{$n=800$}\tabularnewline
\hline 
% JDR & 0.40 & 0.25 & 0.10 & 0.00 &  & 0.80 & 0.50 & 0.80 & 0.80\tabularnewline
MR & 0.06 & 0.08 & 0.11 & 0.12 &  & 0.01 & 0.04 & 0.01 & 0.01\tabularnewline
FAR & 0.17 & 0.10 & 0.11 & 0.10 &  & 0.12 & 0.04 & 0.12 & 0.11\tabularnewline
R1S & 1.00 & 1.00 & 0.91 & 0.92 &  & 1.00 & 1.00 & 0.90 & 0.91\tabularnewline
RCI & 0.91 & 0.87 & 0.91 & 0.91 &  & 0.86 & 0.84 & 0.87 & 0.87\tabularnewline
MSE{*} & 2.45 & 1.77 & 2.95 & 2.50 &  & 2.63 & 2.55 & 3.14 & 2.58\tabularnewline
CV-E & 1.76 & 1.66 & 1.79 & 1.52 &  & 2.04 & 1.66 & 2.46 & 1.94\tabularnewline
\hline 
$p=820$ & \multicolumn{4}{c}{$n=400$} &  & \multicolumn{4}{c}{$n=800$}\tabularnewline
\hline 
% JDR & 0.40 & 0.20 & 0.05 & 0.05 &  & 0.65 & 0.50 & 0.55 & 0.55\tabularnewline
MR & 0.05 & 0.07 & 0.14 & 0.15 &  & 0.02 & 0.04 & 0.03 & 0.03\tabularnewline
FAR & 0.14 & 0.08 & 0.07 & 0.06 &  & 0.10 & 0.05 & 0.08 & 0.08\tabularnewline
R1S & 1.00 & 1.00 & 0.94 & 0.94 &  & 1.00 & 1.00 & 0.93 & 0.93\tabularnewline
RCI & 0.93 & 0.90 & 0.92 & 0.92 &  & 0.88 & 0.86 & 0.89 & 0.89\tabularnewline
MSE{*} & 1.45 & 0.98 & 1.78 & 1.57 &  & 1.19 & 0.92 & 1.47 & 1.21\tabularnewline
CV-E & 1.79 & 1.67 & 1.84 & 1.48 &  & 1.70 & 1.67 & 1.73 & 1.53\tabularnewline
\hline 
\end{tabular}
\caption{Simulation results of Section 5.2. In the tuning process,
\textquotedblleft lambda.min\textquotedblright{} is used. Results are averaged over 20 independent replications. CoxL: unstructured $\ell_1$ penalty (\texttt{glmnet} with \texttt{"cox"} family); sox: our method; .db: with additional debiasing procedure. JDR: joint detection rate, MR: missing rate, FAR: false alarm rate, R1S: rule 1 satisfaction, RCI: the C index of the model with the selected variables, MSE: mean-squared error (*values are multiplied by $10^{-3}$), CV-E: cross-validated error.}
\label{tab:Simulation-high-d}
\end{table}

\begin{table}
\begin{centering}
\begin{tabular}{cccccccccc}
 &  &  &  &  &  &  &  &  & \tabularnewline
\hline 
\hline 
 & \texttt{sox} & \texttt{SGL} & \texttt{grpCox} & \texttt{sox} & \texttt{SGL} & \texttt{grpCox} & \texttt{sox} & \texttt{SGL} & \texttt{grpCox}\tabularnewline
\cline{2-10} \cline{3-10} \cline{4-10} \cline{5-10} \cline{6-10} \cline{7-10} \cline{8-10} \cline{9-10} \cline{10-10} 
 & \multicolumn{3}{c}{Case 1} & \multicolumn{3}{c}{Case 2} & \multicolumn{3}{c}{Case 3}\tabularnewline
\hline 
% JDR & 0.35 & 0.35 & 0.00 & 0.10 & 0.20 & 0.00 & 0.00 & 0.00 & 0.00\tabularnewline
MR & 0.27 & 0.31 & 0.71 & 0.34 & 0.30 & 0.79 & 0.36 & 0.46 & 0.82\tabularnewline
FAR & 0.13 & 0.12 & 0.02 & 0.17 & 0.18 & 0.02 & 0.20 & 0.16 & 0.02\tabularnewline
MSE{*} & 2.04 & 2.02 & 2.21 & 4.38 & 4.19 & 4.77 & 7.28 & 7.25 & 7.49\tabularnewline
\hline 
\end{tabular}
\par\end{centering}
\begin{centering}
MSE: $\|\boldsymbol{\beta}-\hat{\boldsymbol{\beta}}\|_{2}^{2}/p$,
values are multiplied by $10^{-3}$.
\par\end{centering}
\caption{Simulation results of Section 5.3. In the tuning process,
\textquotedblleft lambda.min\textquotedblright{} is used. Results are
averaged over 20 independent replications. \texttt{sox}: our method, \texttt{SGL} the sparse group LASSO, \texttt{grpCox}: the latent overlapping group LASSO. \label{tab:Simulation-nested}}
\end{table}

\begin{table}[!p
]
    \centering
    \begin{tabular}{p{1cm}p{13cm}}
    \hline\hline
    \# & \textbf{Selection rule} \\
    \hline
    1  & If Apixaban is selected, then select DOAC \\
    2  & If Dabigatran is selected, then select DOAC \\
    3  & If DOAC is selected, then select OAC \\
    4  & If High-dose-DOAC is selected, then select DOAC \\
    5...8 & If the interaction of DOAC and a concomitant medication is selected, then both DOAC and the concomitant medication are selected.\\
    \hline\\
    \end{tabular}
    \caption{Selection rules for the case study}
    \label{tab:SelectionRule}
\end{table}

\begin{table}[!p]
    \centering
    \begin{tabular}{cccc}\hline\hline
\textbf{Covariate} & \texttt{sox.refit}&\texttt{CoxL.refit} & \texttt{Cox}\\
\hline
C-Index &0.8034 & 0.7991 &0.8077\\\hline
Age ($\geqslant 75$) &1.84 & 1.80  &1.85  (1.70, 2.03)\\
Sex(female/male) & -&- &0.96  (0.90, 1.03)\\
\multicolumn{4}{c}{\textbf{Comorbidities/Medical score}}\\
CHA$_{2}$DS$_{2}$VAS$_{c}$  ($\geqslant 3$) &0.84  &- &0.90  (0.80, 1.02)\\
Diabetes &- & -&1.08  (1.01, 1.15)\\
COPD/asthma & 1.51  & 1.49  &1.49  (1.41, 1.59)\\
Hypertension &- &- &0.89  (0.81, 0.97)\\
Malignant cancer & 1.56  & 1.58  &1.55  (1.46, 1.65)\\
Stroke & - &- &0.95  (0.88, 1.02)\\
Chronic kidney disease &2.40  & 2.34& 2.39  (2.23, 2.55)\\
Heart disease  &2.55 & 2.36  &2.56  (2.31, 2.83)\\
Major bleeding &1.71  &1.69 &1.72  (1.61, 1.83)\\
\multicolumn{4}{c}{\textbf{OAC use}}\\
DOAC   &0.89 &- &1.04  (0.76, 1.44)\\
Apixaban &0.94  &0.94  & 0.86  (0.67, 1.12)\\
Dabigatran & -&- &0.80  (0.55, 1.17)\\
OAC & 0.17 & 0.16  & 0.17  (0.15, 0.19)\\
High-dose-DOAC &0.91   &- &0.87  (0.69, 1.11)\\
\multicolumn{4}{c}{\textbf{Concomitant medication use}}\\
Antiplatelets &- &- &1.10  (1.03, 1.19)\\
NSAIDs &- &- &1.58  (1.34, 1.86)\\
Statin &0.65  &- &0.63  (0.59, 0.67)\\
Beta-Blockers & 1.04 &- &1.03  (0.97, 1.10)\\
\multicolumn{4}{c}{\textbf{Potential drug-drug interaction}}\\
DOAC: Antiplatelets &- &- &0.67  (0.50, 0.90)\\
DOAC: NSAIDs &- &- & 0.76  (0.43, 1.37)\\
DOAC: Statin &1.05 &0.66   &1.15  (0.90, 1.48)\\
DOAC: Beta-Blockers &0.83 &- &0.86  (0.68, 1.10)\\
\hline\\
    \end{tabular}
    \caption{The estimated hazard ratios and 95\% confidence intervals of each covariate  from various methods. \texttt{CoxL.refit}: unstructured $\ell_1$ penalty; \texttt{sox.refit}: our method; \texttt{Cox}: the standard time dependent-Cox model with all covariates included; C-Index: concordance index}
    \label{tab:ApplicationResults}
\end{table}

\clearpage
\appendix
{\Huge \textbf{Supplementary materials}}
\section{Examples of various types of selection rules and grouping structures}
\label{A:examples}
\begin{example}[Coefficients interpretability] \label{ex:dose}
Consider a cohort study in which patients intermittently receive a drug (treatment) at different dose levels.  The objective is to investigate the association between dose level and time-to-event. Let $X_{1}(t)$ and $X_{2}(t)$ represent indicators for the patient receiving treatment (either high or low dose) at time $t$ and receiving the high dose treatment, respectively. A crucial selection rule in this scenario is 
``if $X_{2}(t)$ is selected, then $X_{1}(t)$ must also be selected''. This is because if $X_{2}(t)$ is selected without $X_{1}(t)$, then the coefficients of $X_{2}(t)$ would be the contrast between taking high-dose treatment versus taking low-dose treatment combined with not taking the treatment, which is not of interest.  Incorporating such selection rules guarantees the interpretability of the selected model's coefficients. To satisfy this selection rule, the grouping structure $\mathbb{G}$ is specified as $\{\{X_{2}(t)\}, \{X_{1}(t), X_{2}(t)\}\}$. It is worth noting that even this simple selection rule necessitates an overlapping grouping structure.
\end{example}

\begin{example}[Strong heredity] 
As mentioned earlier, caution is required when dealing with interaction terms in variable selection. Consider a study involving covariates $X_{1}(t), X_{2}(t)$ and their interaction $X_{3}(t)=X_{1}(t)\times X_{2}(t)$. The strong heredity \citep{lim2015learning} states that ``If the interaction term is selected, then all its main terms must also be selected''. We specify the grouping structure as $\mathbb{G}=\{\{X_{3}(t)\},\{X_{1}(t), X_{3}(t)\},\{X_{2}(t), X_{3}(t)\}\}$.
Incorporating such selection rules can enhance the interpretability of the model, improve statistical power, and simplify experimental designs \citep{Haris_2016}.  In our simulation studies, we present additional examples of grouping structures for more complex scenarios. For instance, in Section \ref{subsec:sim1}, we demonstrate the grouping structure for the case involving interactions between two categorical variables and a continuous variable. In Section \ref{subsec:sim2}, we showcase the strategy for setting the grouping structure when high-dimensional main terms and interactions are included.
\end{example}
% Continuing section \ref{sec:TDCox}, consider the Cox model with both time-dependent covariates and coefficients 
% \begin{equation*}
% h_{i}(t)=h_{0}(t)\cdot\exp\{\mathbf{X}_{i}(t)^{\top}\boldsymbol{\beta}(t)\},  \end{equation*} 
% where $\boldsymbol{\beta}(t)=(\beta_{1}(t),\ldots, \beta_{p}(t))^\top$ with $\beta_{j}(t)=\sum_{m=1}^{M}\theta_{jm}\phi_{m}(t)$ for $\ j=1,\ldots,p$. 
% The problem of estimating $\beta_{j}(t)$ for $j=1, \ldots, p$ is then transformed
% to the problem of estimating $\boldsymbol{\theta}_{j}=(\theta_{j1},\ldots,\theta_{jM})^{\top}$ for $j=1, \ldots, p$. Let $\boldsymbol{\Theta}=(\boldsymbol{\theta}_{1},\ldots,\boldsymbol{\theta}_{p})^{\top}$
% be the $p\times M$ matrix of parameters to be estimated and $\boldsymbol{\phi}(t)=(\phi_{1}(t),\ldots,\phi_{M}(t))^{\top}$. Thus $\boldsymbol{\beta}(t)=\mathbf{\boldsymbol{\Theta}}\boldsymbol{\phi}(t)$, which gives
% \begin{equation*}
% h_{i}(t)=h_{0}(t)\cdot\exp\{\mathbf{X}_{i}(t)^{\top}\boldsymbol{\Theta}\boldsymbol{\phi}(t)\}. 
% \end{equation*}
% Suppose the selection rule to be respected is ``all the basis functions related to one variable should be selected collectively'', or ``$ \mathbb{g}_{j} = \{\theta_{jm}\}_{m=1}^{M}$  should be selected collectively for any $j=1, \ldots, p$''. Define $\mathbb{V} = \{\theta_{jm}, j=1, \dots, p; m=1\dots,M\}$ and $\mathbb{g}=\{\{\theta_{jm}\}_{m=1}^{M}, j=1\dots,p\}$. Then the selection can be respected.
\begin{example}[Incorporating temporal structure]\label{ex:TDcoef}
In studies involving patients with dementia, it is common to categorize them into four phases: mild, moderate, moderately severe, and severe cognitive decline \citep{reisberg1982global}.  Consider a cohort study focusing on patients diagnosed with mild cognitive decline and examining the association between blood pressure $X(t)$ at each phase and the time to severe cognitive decline. We define $T_1$ as the time of diagnosis for moderate cognitive decline and $T_2$ as the time for moderately severe cognitive decline. To capture the temporal aspect, we include the time-dependent covariates  $Z_{1}(t)=I(t<T_{1})X(t)$, $Z_{2}(t)=I(T_{1}\leqslant t<T_{2})X(t)$ and $Z_{3}(t)=I(t \geqslant T_{2})X(t)$ in the Cox model. Suppose we have a priori knowledge that if the blood pressure in a previous phase is associated with the outcome, the blood pressure in the later phase should also be associated. This implies the following selection rules: ``if $Z_1(t)$ is selected, then $Z_2(t)$ and $Z_3(t)$ should also be selected" and ``if $Z_2(t)$ is selected, then $Z_3(t)$ should be selected". By incorporating such selection rules, we can accommodate the temporal structure of the time-dependent covariate. The corresponding grouping structure is $\mathbb{G}=\{\{Z_{1}(t)\}, \{Z_{1}(t), Z_{2}(t)\}, \{Z_{1}(t), Z_{2}(t), Z_{3}(t)\}\}$. This example illustrates the use of step functions to model time-dependent associations. 

More flexibly, consider the Cox model with both time-dependent covariates and coefficients, 
$
h\{t|\mathbf{X}(t)\}=h_{0}(t)\cdot\exp\{\mathbf{X}(t)^{\top}\boldsymbol{\beta}(t)\}, 
$
where $\boldsymbol{\beta}(t)=\{\beta_{1}(t),\ldots, \beta_{j}(t)\}^\top$ with $\beta_{j}(t)=\sum_{m=1}^{M}\theta_{j,m}\phi_{m}(t)$, for $\ j=1,\ldots,p$. 
Rewriting, we have:
$
h\{t|\mathbf{Z}(t)\}=h_{0}(t)\cdot\exp\{\mathbf{Z}(t)^{\top}\boldsymbol{\theta}\}
$,
where $Z_{j,m}(t)=X_j(t)\phi_{m}(t)$, e.g.,
\begin{equation*}
\mathbf{Z}(t)=
(\underbrace{X_1(t)\phi_{1}(t), \dots, X_1(t)\phi_{M}(t)}_{\mathbf{Z}_{1,\cdot}(t)},
\dots,
\underbrace{X_p(t)\phi_{1}(t), \dots, X_p(t)\phi_{M}(t)}_{\mathbf{Z}_{p,\cdot}(t)})^{\top}
=(\mathbf{Z}_{1,\cdot}(t),\dots,\mathbf{Z}_{p,\cdot}(t))^{\top},
\end{equation*}
which has $p\times M$ elements. Thus, in the context of predictor identification, selecting $X_j$ is equivalent to selecting $\mathbf{Z}_{j,\cdot}(t)$. In the context of estimation, estimating $\boldsymbol{\beta}(t)$ equates to estimating the ${p\times M}$ length vector $\boldsymbol{\theta}=(\boldsymbol{\theta}_{1,\cdot},\dots,\boldsymbol{\theta}_{p,\cdot})$, where $\boldsymbol{\theta}_{j,\cdot}=\{\theta_{j,1},\dots,\theta_{j,M}\}$. Suppose the selection rule to be respected is "all the basis functions related to one variable should be selected collectively" or ``$\mathbf{Z}_{j,\cdot}(t)=\{Z_{j,1}(t),\dots,Z_{j,M}(t)\}, \forall j=1,\dots,p$ should be selected collectively''. By incorporating such selection rules, we can avoid the selection discrepancy among basis functions while accommodating the time-dependent coefficient feature in Cox models.  We can define $\mathbb{V}=\{Z_{j,m}(t), j=1,\dots,p, m=1\dots,M\}$ and $\mathbb{G}=\{\mathbf{Z}_{j\cdot}(t), j=1\dots,p\}$, and the selection rule can be respected.
\end{example}
\begin{example}[Incorporating spatial structure] Suppose that high-dimensional voxel signals  over time in patients' brains are collected in functional magnetic resonance imaging (fMRI) data. The goal is to identify the regions (of voxels) and individual voxels associated with the time to cocaine relapse \citep{zhai2021cox} or Alzheimer's disease progression \citep{lee2016sparse}. Variable selection conducted to analyze such data should be informed by the three-dimensional grid structure (such as localized clusters on the brain) \citep{chklovskii2004maps}. Here we take a simplified example for illustration. Suppose there are three contiguous voxels; the intensities of the signals are denoted  $X_1(t)$, $X_2(t)$, and $X_3(t)$. Consider the case where two parcels (clusters) $\{X_1(t), X_2(t)\}$ and $\{X_1(t), X_2(t), X_3(t)\}$ are hierarchically constructed \citep{ward1963hierarchical}, in which only neighboring voxels can be merged together. Larger parcels can be regarded as potential regions of interest. To incorporate such a structure, we first define $X_4(t)$ and $X_5(t)$ as the average of the two parcels. Then we set selection rules as ``if either $X_1(t)$ or $X_2(t)$ is selected, then select $X_4(t)$'', and ``if either $X_3(t)$ or $X_4(t)$ is selected, then select $X_5(t)$'' to encode the hierarchy and promote the parcel selection \citep{jenatton2012multiscale}.  The corresponding $\mathbb{G}$ is $\{\{X_1(t)\}, \{X_2(t)\}\{X_3(t)\}, \{X_1(t), X_2(t), X_4(t)\}, \{X_1(t), X_2(t), X_3(t), X_4(t), X_5(t)\}\}$. The above example represents the case where two parcels are nested. Incorporating more complex structures, such as multiple overlapped parcels nesting in different larger parcels, is also possible.
\end{example}

\begin{example}[Tree and directed acyclic graph grouping structures]

In certain real-world scenarios, more complex structures of covariates, such as trees \citep{jenatton2011proximal} and directed acyclic graphs \citep{jenatton2011structured}, can be incorporated into variable selection. These structures allow for more intricate relationships among variables to be taken into account. For example, in our model, covariates can be represented as nodes in a tree, and users can specify that a variable is selected only if all its ancestors in the tree are already selected. Moreover, the framework can be further extended to include directed cyclic graphs, which have been found useful in hierarchical variable selection. These enhancements provide additional flexibility in capturing complex relationships among variables.
\end{example}

\section{Steps of \texttt{computeFlow}}\label{A:BasicSteps}
Table \ref{tab:BasicSteps} shows the steps of \texttt{computeFlow}.
\begin{table}[h]
    \centering
    \begin{tabular}{p{3cm}p{12cm}}
    \hline
     \textbf{Step} & \textbf{Details}\\\hline
    Projection step & solve a relaxed version of (\ref{eq:dual}) to calculate $\boldsymbol{\gamma}$, which is the lower bound of $\frac{1}{2t}\left\| \left\{ \tilde{\boldsymbol{\beta}}-t\nabla f(\tilde{\boldsymbol{\beta}})  \right\} -\boldsymbol{\gamma} \right\|_{2}^{2}$, and also satisfies $\sum_{j}\gamma_{j}\leqslant\lambda\sum_{\mathbb{g}\in\mathbb{G}}\omega_{|\mathbb{g}}$.
    The value of $\boldsymbol{\gamma}$ is the projection of the vectors $\boldsymbol{\xi}_{|\mathbb{g}}$.\\\\
     Updating step  &  update $(\sum_{\mathbb{g}\in\mathbb{G}}\boldsymbol{\xi}^j_{|\mathbb{g}})_{X_j\in\mathbb{V}}$ by maximizing $\sum_{X_j\in\mathbb{V}}\sum_{\mathbb{g}\in\mathbb{G}}\boldsymbol{\xi}^j_{|\mathbb{g}}$, while keeping $\sum_{X_j\in\mathbb{g}}\boldsymbol{\xi}^j_{|\mathbb{g}}\leqslant\lambda\omega_{\mathbb{g}}$. By doing so, we can ensure that the constraint in (\ref{eq:dual}) holds. This can be done by the max flow algorithm. Details of the implementation can be found in Section \ref{subsec:masflow}.\\\\ 
     Recursion step/divide and conquer & According to the mini-cut theorems \citep{ford1956maximal}, define $\mathbb{V}^{*}=\{X_j\in\mathbb{V}: \sum_{\mathbb{g}\in\mathbb{G}}\boldsymbol{\xi}^j_{|\mathbb{g}}=\gamma_j\}$, and $\mathbb{G}^{*}=\{\mathbb{g}\in\mathbb{G}: \sum_{X_j\in\mathbb{g}}\boldsymbol{\xi}^j_{|\mathbb{g}}<\lambda\omega_{\mathbb{g}}\}$. Then apply steps 1 and 2 to $(\mathbb{V}^{*}, \mathbb{G}^{*})$ and their respective complements until $(\sum_{\mathbb{g}\in\mathbb{G}}\boldsymbol{\xi}^j_{|\mathbb{g}})_{X_j\in\mathbb{V}}$ (obtained from step 2) matches $\boldsymbol{\gamma}$ (obtained from step 1).\\
    \hline\\
    \end{tabular}
    \caption{Steps of \texttt{computeFlow}}
    \label{tab:BasicSteps}
\end{table}
\section{Grouping structure identification in the simulation}
\label{A:GSI}
The developed methods (similar to the overlapping group Lasso in \citep{wang2021general}) can enforce a number of groups of variable coefficients to be 0 with a certain level of penalization. The remaining variables are said to be selected.\\
For the ease of notation, we use $A$ to represent $A(t)$. Consider the selection rule ``if $\{A_{1}B, A_{2}B\}$ is selected, then $\{A_{1}, A_{2}, B\}$ must be selected''. Suppose for now all candidate variables are $\mathbb{V}=\{A_{1}, A_{2}, B, A_{1}B, A_{2}B\}$, which are the variables that are involved in this rule. According to Table 2 in \citep{wang2021general}, the selection dictionary (all permissible subsets of covariates that respect the selection dependency) is $$
\mathbb{D}=\left\{\emptyset, \{A_{1}, A_{2}\}, \{B\}, \{A_{1}, A_{2}, B\}, \{A_{1}, A_{2}, B, A_{1}B, A_{2}B\}\right\}.
$$
Based on Theorem 5 in \citep{wang2021general}, we need to create groups whose complements (and their combinations) are equal to $\mathbb{D}\setminus\mathbb{V}$. We thus postulate three groups: $\{A_{1}, A_{2}, A_{1}B, A_{2}B\} \mathrel{,} \{B, A_{1}B, A_{2}B\}$ and $\{A_{1}B, A_{2}B\}$, which satisfy the requirement. Similarly, to respect the selection dependency ``if $\{C_{1}B, C_{2}B\}$ is selected, then $\{C_{1}, C_{2}, B\}$ must be selected'', we postulate another three groups $\{C_{1}, C_{2}\mathrel{,} C_{1}B\mathrel{,} C_{2}B\} \mathrel{,} \{B, C_{1}B, C_{2}B\}$ and $\{C_{1}B, C_{2}B\}$. \\
However, the two rules share a same variable $B$: if either $\{A_{1}B, A_{2}B\}$ or $\{C_{1}B, C_{2}B\}$ is selected, then $B$ must be selected. To satisfy this requirement, we need to merge the two groups $\{B, A_{1}B, A_{2}B\}$ and $\{B, C_{1}B, C_{2}B\}$ into one group $\{A_{1}B\mathrel{,} A_{2}B\mathrel{,} B\mathrel{,} C_{1}B\mathrel{,} C_{2}B\}$ to prevent the occurrence of rule-breaking combinations for example, $\{C_{1}B, C_{2}B\}$ being selected without $B$.\\
We also need to respect another selection rule: the dummy variables for a categorical variable need to be selected collectively. The categorical interaction variables $AB$ and $BC$ are already being selected collectively because of the above selection dependencies. However, additional groups for $\{A_{1}, A_{2}\}$ are unnecessary as this would make it possible to select $\{A_{1}B, A_{2}B\}$  without $A$. In addition, with the above groups, $A_{1}, \text{ and } A_{2}$ would never be selected individually because they are always in a same group.\\
Therefore, we have 5 defined groups listed below \begin{align*}
 \mathbb{g}_1 = \{A_{1}, A_{2}, A_{1}B, A_{2}B\},
\mathbb{g}_2  = \{B, A_{1}B, A_{2}B, C_{1}B, C_{2}B\},\\
\mathbb{g}_3  = \{A_{1}B, A_{2}B\},
\mathbb{g}_4  = \{C_1, C_2, C_{1}B, C_{2}B\},
\mathbb{g}_5  = \{C_{1}B, C_{2}B\}.
\end{align*} 
Our resulting selection dictionary is: 
$\{\emptyset\mathrel{,}
\{B\}\mathrel{,}
\{C_{1}, C_{2}\}\mathrel{,}
\{B, C_{1}, C_{2}\}\mathrel{,}
\{B, C_{1}, C_{2}, C_{1}B, C_{2}B\}\mathrel{,}
\{A_{1}, A_{2}\}\mathrel{,}
\{A_{1}B, A_{2}B, B\}\mathrel{,}
\{A_{1}, A_{2}, C_{1}\mathrel{,} C_{2}\}\mathrel{,}
\{A_{1}, A_{2}, A_{1}B, A_{2}B, B\}\mathrel{,}
\{A_{1}, A_{2}, B, C_{1}, C_{2}\}\mathrel{,}
\{A_{1}, A_{2}, B, C_{1}, C_{2}, A_{1}B, A_{2}B\}\mathrel{,}
\{A_{1}, A_{2}, B\mathrel{,} C_{1}, C_{2}, C_{1}B, C_{2}B\}\mathrel{,}
\{A_{1}, A_{2}\mathrel{,} B, A_{1}B, A_{2}B \mathrel{,}C_{1}, C_{2}, C_{1}B,  C_{2}B\}\}$. 
The code to derived the selection dictionary using \texttt{R} is available at \url{https://github.com/Guanbo-W/sox_sim}. One can verify the correctness of the derived selection dictionary using Theorem 5 in \citep{wang2021general}.

\section{Additional simulation results for Section 5.2}\label{app:add_5.2}
See the results in Table \ref{tab:Simulation-high-d-1se}
\begin{table}
\begin{centering}
\begin{tabular}{cccccccccc}
 &  &  &  &  &  &  &  &  & \tabularnewline
\hline 
\hline 
Method & sox & sox.db & CoxL & CoxL.db &  & sox & sox.db & CoxL & CoxL.db\tabularnewline
\hline 
$p=210$ & \multicolumn{4}{c}{$n=400$} &  & \multicolumn{4}{c}{$n=800$}\tabularnewline
\hline 
JDR & 0.15 & 0.05 & 0.00 & 0.00 &  & 0.45 & 0.00 & 0.40 & 0.05\tabularnewline
MR & 0.17 & 0.28 & 0.30 & 0.37 &  & 0.05 & 0.15 & 0.05 & 0.13\tabularnewline
FAR & 0.04 & 0.00 & 0.02 & 0.01 &  & 0.02 & 0.00 & 0.03 & 0.01\tabularnewline
R1S & 1.00 & 1.00 & 0.98 & 0.99 &  & 1.00 & 1.00 & 0.98 & 0.99\tabularnewline
RCI & 0.83 & 0.81 & 0.82 & 0.81 &  & 0.83 & 0.82 & 0.83 & 0.82\tabularnewline
MSE{*} & 7.01 & 6.39 & 9.17 & 6.93 &  & 5.79 & 5.76 & 6.48 & 5.70\tabularnewline
CV-E & 1.82 & 1.78 & 1.87 & 1.77 &  & 1.73 & 1.73 & 1.75 & 1.71\tabularnewline
\hline 
$p=465$ & \multicolumn{4}{c}{$n=400$} &  & \multicolumn{4}{c}{$n=800$}\tabularnewline
\hline 
MR & 0.18 & 0.33 & 0.36 & 0.40 &  & 0.04 & 0.13 & 0.07 & 0.13\tabularnewline
FAR & 0.02 & 0.00 & 0.01 & 0.01 &  & 0.01 & 0.00 & 0.01 & 0.00\tabularnewline
R1S & 1.00 & 1.00 & 0.99 & 0.99 &  & 1.00 & 1.00 & 0.99 & 0.99\tabularnewline
RCI & 0.84 & 0.81 & 0.83 & 0.81 &  & 0.83 & 0.82 & 0.83 & 0.82\tabularnewline
MSE{*} & 3.22 & 2.85 & 3.71 & 3.00 &  & 2.63 & 2.55 & 3.14 & 2.58\tabularnewline
CV-E & 1.83 & 1.78 & 1.87 & 1.74 &  & 1.71 & 1.71 & 1.76 & 1.70\tabularnewline
\hline 
$p=820$ & \multicolumn{4}{c}{$n=400$} &  & \multicolumn{4}{c}{$n=800$}\tabularnewline
\hline 
MR & 0.20 & 0.31 & 0.40 & 0.45 &  & 0.04 & 0.18 & 0.10 & 0.16\tabularnewline
FAR & 0.02 & 0.00 & 0.01 & 0.01 &  & 0.01 & 0.00 & 0.01 & 0.00\tabularnewline
R1S & 1.00 & 1.00 & 0.99 & 0.99 &  & 1.00 & 1.00 & 0.99 & 0.99\tabularnewline
RCI & 0.84 & 0.80 & 0.82 & 0.81 &  & 0.83 & 0.82 & 0.84 & 0.83\tabularnewline
MSE{*} & 1.89 & 1.69 & 2.27 & 1.87 &  & 1.49 & 1.46 & 1.78 & 1.40\tabularnewline
CV-E & 1.82 & 1.78 & 1.90 & 1.74 &  & 1.73 & 1.73 & 1.78 & 1.69\tabularnewline
\hline 
\end{tabular}
\par\end{centering}
\begin{centering}
\par\end{centering}
\caption{Simulation results of the high-dimensional case. In the tuning process,
\textquotedblleft lambda.1se\textquotedblright{} is used. Results are averaged over 20 independent replications. CoxL: unstructured $\ell_1$ penalty (\texttt{glmnet} with \texttt{"cox"} family); sox: our method; .db: with additional debiasing procedure. JDR: joint detection rate, MR: missing rate, FAR: false alarm rate, R1S: rule 1 satisfaction, RCI: the C index of the model with the selected variables, MSE: mean-squared error (*values are multiplied by $10^{-3}$), CV-E: cross-validated error. \label{tab:Simulation-high-d-1se}}
\end{table}
{
\section{Sample solution path}\label{app:samplepath}
See Figure \ref{fig:Sample-solution-paths} for the sample solution path.
\begin{figure}
\begin{centering}
\includegraphics[scale=0.8]{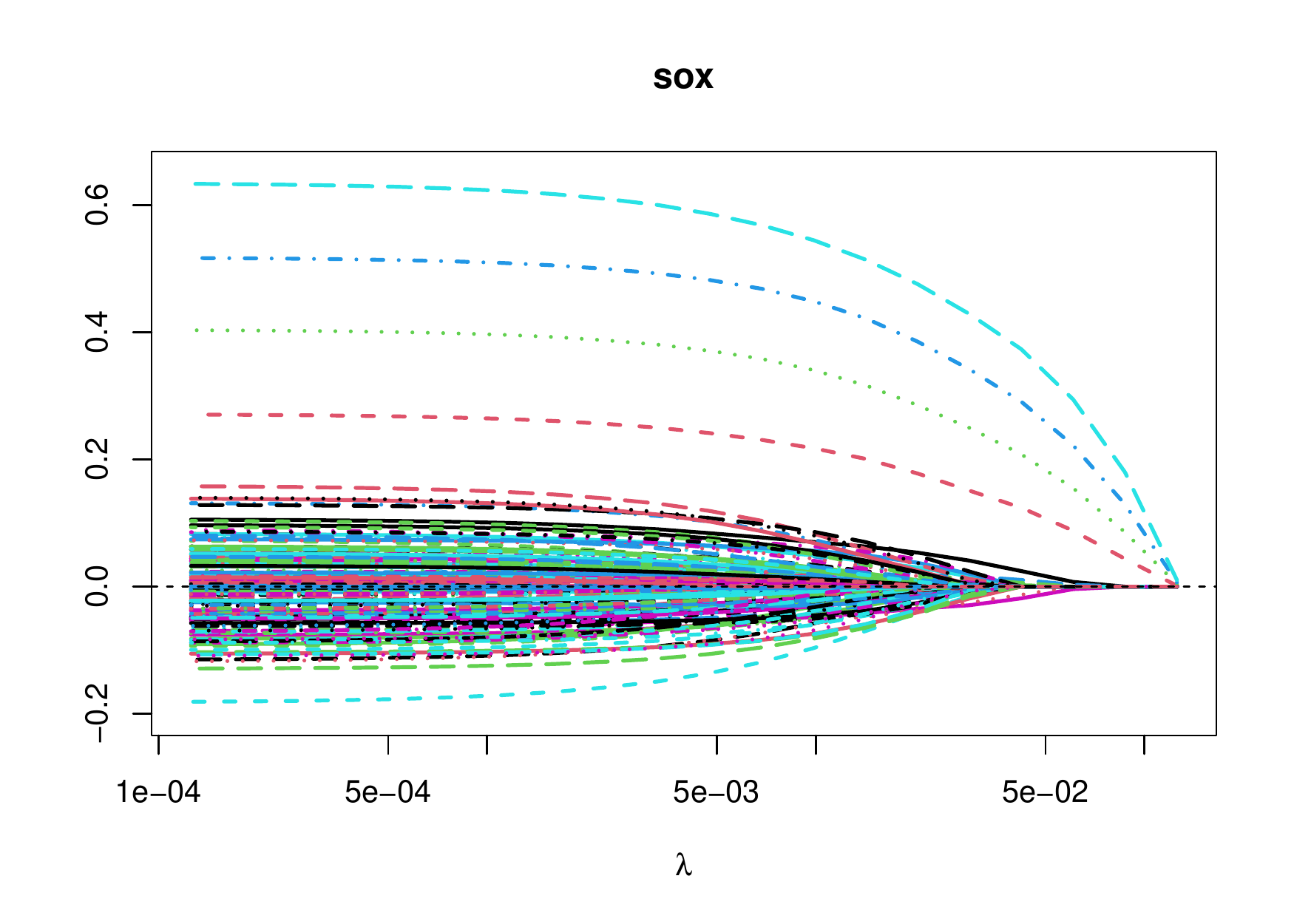}
\par\end{centering}
\begin{centering}
\includegraphics[scale=0.8]{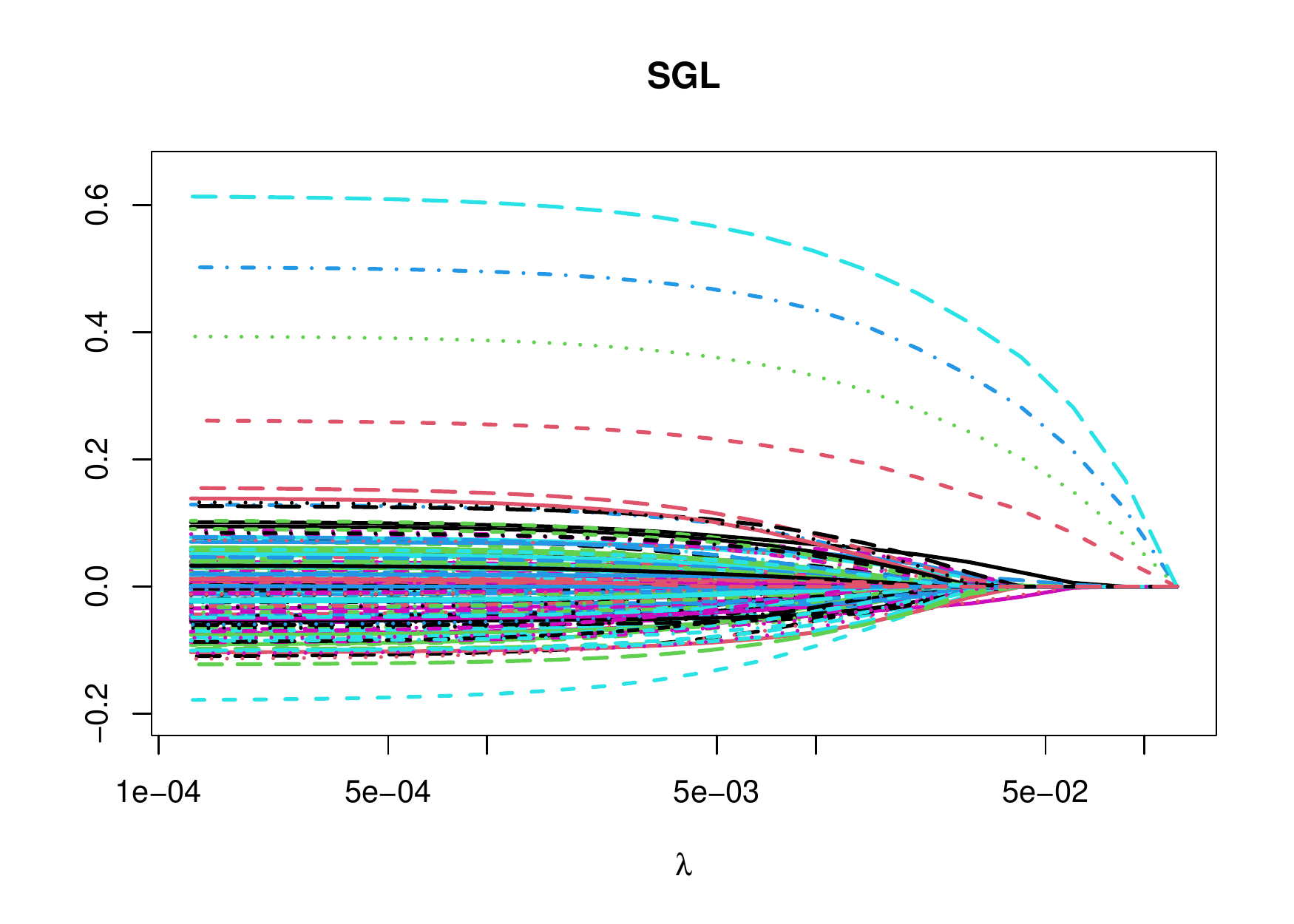}
\par\end{centering}
\caption{Sample solution paths from \texttt{sox} and \texttt{SGL} using the same data and the
same $\lambda$ sequence.\label{fig:Sample-solution-paths}}
\end{figure}

\section{Additional simulation: Comparison with existing sparse group lasso methods (additional simulations)}\label{app:5.3add}
In this simulation, we compare our method, the sparse group LASSO, and the LASSO, implemented by \texttt{sox}, \texttt{SGL}, and \texttt{glmnet} respectively. We aim to show that \texttt{sox} can respect certain selection rules that \texttt{SGL} or \texttt{glmnet} cannot. We follow a similar setting as the one in Section 5.2 within the cases of $(n=400/800, p=210)$. Since \texttt{SGL} cannot handle time-dependent Cox models, we generate time-fixed covariates.

The sparse group LASSO, due to its inability to accommodate overlapping groups, does not adhere to the principle of strong heredity. To define its group structure, we have organized the groups in a specific manner: each of the 10 groups contains two consecutive main terms along with their interaction (we have 20 main terms, so 10 such groups are specified, each containing three variables). Additionally, the 180 remaining interactions are each treated as individual groups. This configuration results in a total of 190 distinct groups.

The results, presented in Table \ref{tab:sim-toSGL}, show that \texttt{sox} outperforms \texttt{glmnet}, consistent with the findings in Section 5.2. The inferior performance of \texttt{glmnet} is attributed to its inability to incorporate selection rules. In this simulation, a flawed grouping structure was applied to the sparse group LASSO, leading it to adhere to selection rules not aligned with the actual data generation mechanism. This situation parallels Bayesian analysis, where an incorrect prior leads to suboptimal outcomes, which explains why \texttt{SGL} demonstrates the least effective performance in this context. \\

\begin{table}
\begin{centering}
\begin{tabular}{cccccccccccc}
\toprule 
Method & sox & sox.db & CoxL & CoxL.db & SGL &  & sox & sox.db & CoxL & CoxL.db & SGL\tabularnewline
\midrule 
$p=210$ & \multicolumn{5}{c}{$n=400$}  &  & \multicolumn{5}{c}{$n=800$} \tabularnewline
\midrule
MR & 0.04 & 0.05 & 0.06 & 0.08 & 0.02 &  & 0.00 & 0.01 & 0.00 & 0.00 & 0.18\tabularnewline
FAR & 0.32 & 0.22 & 0.17 & 0.14 & 0.76 &  & 0.34 & 0.24 & 0.22 & 0.18 & 0.40\tabularnewline
R1S & 1.00 & 1.00 & 0.89 & 0.91 & 0.81 &  & 1.00 & 1.00 & 0.87 & 0.90 & 0.87\tabularnewline
RCI & 0.84 & 0.84 & 0.84 & 0.84 & 0.88 &  & 0.82 & 0.82 & 0.82 & 0.82 & 0.72\tabularnewline
MSE{*} & 3.41 & 2.37 & 3.77 & 3.18 & 3727 &  & 2.21 & 1.50 & 2.37 & 1.80 & 3242\tabularnewline
CV-E & 5.51 & 5.34 & 5.52 & 5.30 & 48.02 &  & 5.48 & 5.38 & 5.48 & 5.36 & 53.14\tabularnewline
\bottomrule
\end{tabular}
\par\end{centering}
\caption{Simulation results of interaction selection under the time-fixed, high-dimensional setting. In the tuning process,
\textquotedblleft lambda.1se\textquotedblright{} is used. Results are averaged over 20 independent replications. CoxL: unstructured $\ell_1$ penalty (\texttt{glmnet} with \texttt{"cox"} family); sox: our method; SGL: the sparse group LASSO; .db: with additional debiasing procedure. JDR: joint detection rate, MR: missing rate, FAR: false alarm rate, R1S: rule 1 satisfaction, RCI: the C index of the model with the selected variables, MSE: mean-squared error (*values are multiplied by $10^{-3}$), CV-E: cross-validated error.}
\label{tab:sim-toSGL}
\end{table}
}
\section{Additional simulation: Timing results.}\label{app:time}

Additionally, we also tested the computational speed of \texttt{sox}. We adopted the simulation setting from Section 5.2. For all $(n,p)$ pairs, we reported the computation time (in seconds) for solving 10-fold cross-validation on the same $\lambda$ sequence of length 30 in Table \ref{tab:Timing-results-highD}. The computation time of \texttt{sox.db} does not include the necessary procedures to acquire the initial estimates (\texttt{sox} in our case) used to calculate the regularization weights. Our findings indicate that for a complex case where the sample size is $n=800$ and the number of variables is $p=820$, a comprehensive analysis can be completed in approximately 10 minutes. In contrast, for a simpler scenario with a sample size of $n=400$ and $p=210$ variables, the analysis requires less than 30 seconds to finish.
\begin{table}
\begin{centering}
\begin{tabular}{ccccc}
\toprule 
 & \multicolumn{2}{c}{$n=400$} & \multicolumn{2}{c}{$n=800$}\tabularnewline
\cmidrule{2-5} \cmidrule{3-5} \cmidrule{4-5} \cmidrule{5-5} 
 & sox & sox.db & sox & sox.db\tabularnewline
$p=210$ & 26.80 & 26.47 & 54.13 & 53.93\tabularnewline
$p=465$ & 121.39 & 114.12 & 217.95 & 217.11\tabularnewline
$p=820$ & 391.26 & 359.85 & 704.67 & 698.86\tabularnewline
\bottomrule
\end{tabular}
\par\end{centering}
\caption{Timing results of the high-dimensional case. Results are averaged over 20 independent replications. \label{tab:Timing-results-highD}}
\end{table}

\section{Additional simulation: Comparison of different group sizes, the amount of overlap, and the sparsity levels}\label{app:add_groupsize}
In this section, we delved into how group size, the amount of overlap between groups, and sparsity level influence the performance of our method. We adopt a low-dimensional setting ($n=400, p=25$) time-dependent (with four time-points, and each variable held constant for two or three times-points).
Each variable is independently generated by the standard Gaussian
distribution. We investigate seven grouping structures with different
group sizes and amount of overlap, the details of which are summarized
in Table \ref{tab:grp-overlap-size-setup}. 
% The effects of groups sizes and overlap sizes are less prominent than the effect of the sparsity of the group. Specifically, The third (group size 10, overlap size 5) and fourth (group size 13, overlap size 5) settings have significantly worse overall performance than the other three settings, which is a result of higher chance to exclude Group 2 (in blue) due to its higher sparsity.

The results of the simulation are detailed in Table \ref{tab:grp-overlap-size-result}. A closer look at settings 1, 3, and 2, which feature groups of 10 variables each with similar sparsity but different overlap sizes (8, 5, and 2, respectively), reveals a decrease in false alarm rates. This trend is attributed to the fact that smaller overlaps reduce the probability of mistakenly selecting variables from both groups, thereby lowering the chances of misidentifying noise as a significant signal. Although setting 1 shows a slightly higher sparsity level, it does not substantially affect the outcome.

In contrast, settings 4, 3, and 5, which have identical overlap and sparsity levels but varying group sizes (13, 10, and 7, respectively), also show a declining trend in false alarm rates. This is because the selection of variables 14 and 15 may result in the selection of the variables in group 2. When group 2 has more variables, there is an increased risk of erroneously selecting them.

Regarding settings 5 to 7, where each group consists of 7 variables with an overlap size of 5 and varying sparsity levels for group 2 (0.7, 0.9, and complete sparsity or 1), settings 5 and 6 display similar false alarm rates. This is due to the possibility of a single signal in a group triggering the erroneous selection of all variables in that group. However, with complete sparsity, the false alarm rate tends to zero. Notably, setting 6 exhibits a significantly higher missing rate, possibly because a lone signal in group 2 (variable 14) is sometimes not strong enough for selection, leading to its omission.

These findings demonstrate that group size, overlap, and sparsity levels significantly influence the performance of \texttt{sox}. Nonetheless, caution should be exercised in generalizing these results, as variations in the data-generating mechanism can alter these conclusions. 

\begin{table}
\begin{centering}
\includegraphics[width=1\textwidth]{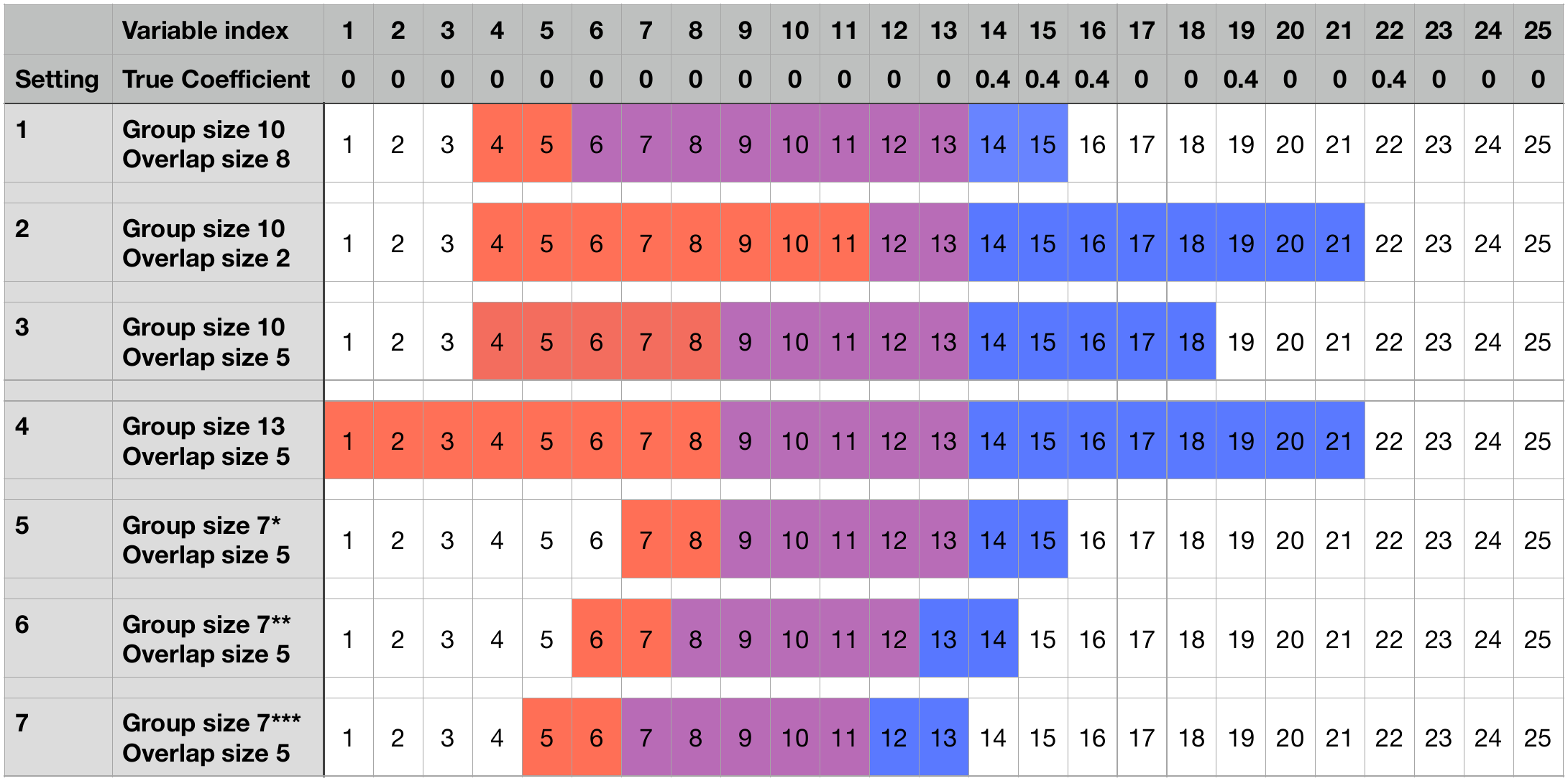}
\par\end{centering}
\caption{True coefficients and grouping structures for the simulation to evaluate
the effects of group sizes and the amount of overlaps. In all seven
settings, there is a group 1 in red and a group 2 in blue with their
overlaps in purple. For the variables that are in neither groups 1
or 2, each belongs to a group of size one.  \label{tab:grp-overlap-size-setup}}

\end{table}

\begin{table}
\begin{centering}
\begin{tabular}{cccccccc}
\toprule 
Setting & 1 & 2 & 3 & 4 & 5 & 6 & 7\tabularnewline
Overlap size &  8 &   2 &   5 & 5 & 5 & 5 & 5\tabularnewline
Group size & 10 & 10 &   10 &   13 &   7 & 7 & 7\tabularnewline
SL* of group 2 & 0.8 & 0.7 &0.7 & 0.7 &   0.7&   0.9 &   1 \tabularnewline
SL* of group 1 & 1 & 1 &1 & 1 & 1 & 1 & 1 \tabularnewline
\midrule
MR & 0.01 & 0.00 & 0.01 & 0.01 & 0.00 & 0.11 & 0.00\tabularnewline
FAR & 0.36 & 0.30 & 0.31 & 0.46 & 0.17 & 0.17 & 0.01\tabularnewline
RCI & 0.76 & 0.76 & 0.76 & 0.76 & 0.76 & 0.75 & 0.76\tabularnewline
MSE{**} & 1.64 & 1.57 & 1.61 & 1.66 & 1.51 & 1.76 & 1.55\tabularnewline
CV-E & 1.90 & 1.89 & 1.89 & 1.89 & 1.88 & 1.91 & 1.88\tabularnewline
\bottomrule
\end{tabular}
\par\end{centering}
\caption{Simulation results for different group sizes and amount of overlap. {*} Sparsity Level; 
{**} MSEs are multiplied by $10^{-2}$.\label{tab:grp-overlap-size-result}}

\end{table}

\section{Inclusion and exclusion criteria}\label{A:exclusion}
\begin{figure}[ht]
    \centering
    \includegraphics{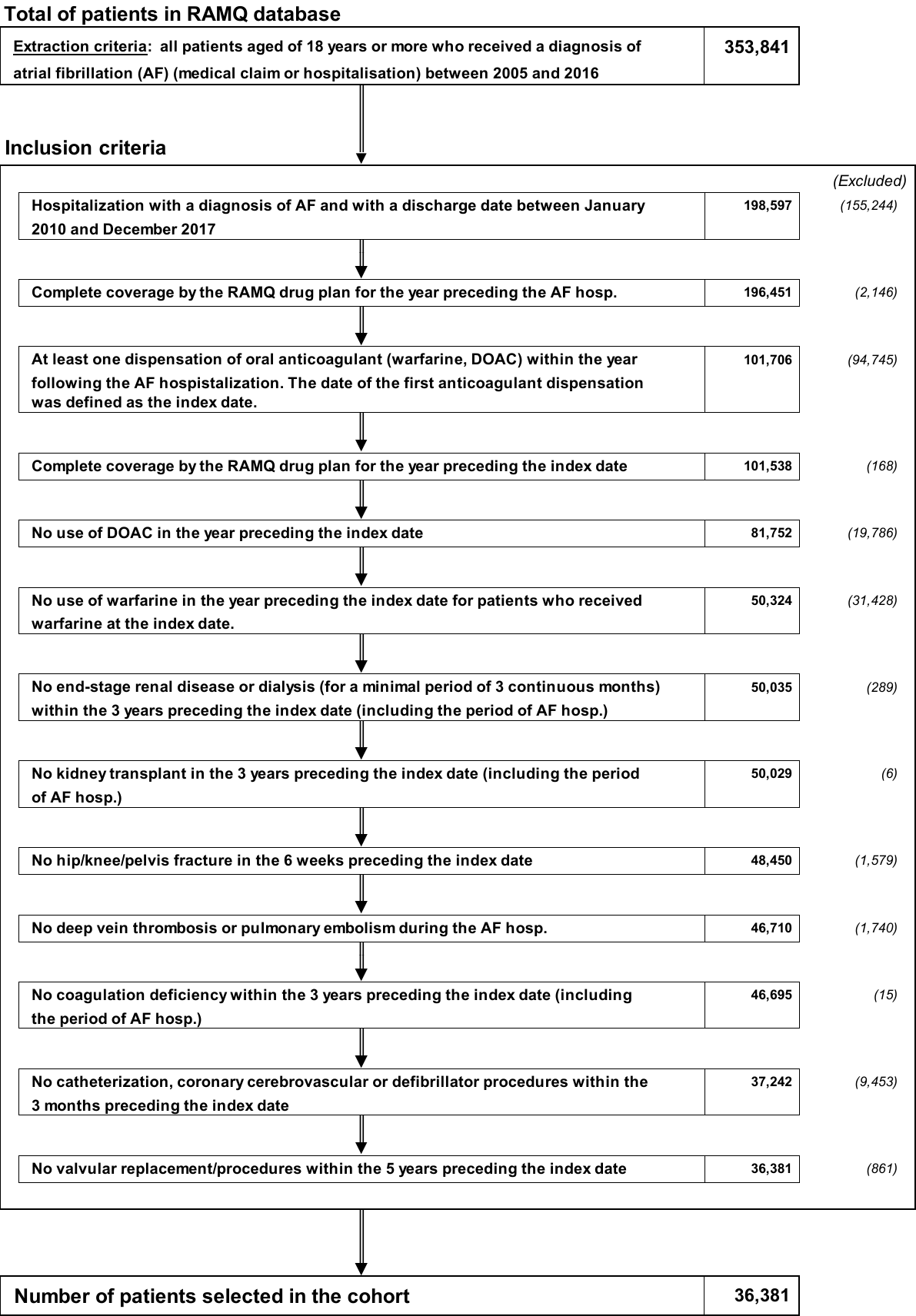}
    \caption{The inclusion and exclusion criteria for the patients of the study cohort. (AF: atrial fibrillation; OAC: oral anticoagulants).}
    \label{fig：inclu/exclu}
\end{figure}

Figure \ref{fig：inclu/exclu} shows the inclusion and exclusion criteria for the patients of the study cohort.

\section{Covariate definitions}
\label{A:CovariateDefinition}
The 7 baseline covariates are 1) Age ($\geqslant 75/<75$), 2) Sex(female/male), 3) CHA2DS2VASc  ($\geqslant 3/<3$), 4) Diabetes, 5) COPD/asthma, 6) Hypertension and 7) Malignant cancer. All other covariates are time-dependent covariates.\\
Heart disease: including 1) valvular heart disease, 2) peripheral vascular disease, 3) cardiovascular disease, 4) chronic heart failure, and 5) myocardial infarction.\\
DOAC: it is 1 if the patient uses DOAC, 0 if the patient is taking warfarin or not taking any OAC at the time $t$.\\
OAC: it is 1 if the patient uses OAC, 0 if the patient is not taking any OAC at the time $t$.\\
High-dose-DOAC: it is 1 if the patient uses high-dose-DOAC, 0 if the patient is taking low-dose-DOAC, or warfarin or not taking any OAC at the time $t$.\\
Antiplatelets: ASA low dose ($\geqslant 80$ mg and $\leqslant 260$ mg), dipyridamole or clopidrogel or ticlopidine or prasugrel or ticagrelor.\\
Statin: Statin or other lipid lowering drugs.\\
NASIDs: Nonsteroidal anti-inflammatory drugs.
\section{Summary statistics of all the covariates}
\label{A:Tableone}
Table \ref{A:tab:Tableone} gives the summary statistics of all the covariates stratified by if the subject experienced the event (death) during the follow-up. All the covariates in this analysis are binary variables. The first two columns (\% at cohort entry) show the summary statistics of the covariates at the time of cohort entry. The second two columns (\% of value changed) show, for each covariate, the percentage of patients who experienced situation change during the follow-up. The third two columns (\% mean over population and time) show, for each covariate, the average value of the covariates across all patients  during the follow-up (the mean, over the population, of the values of the covariates during the follow-up).
\begin{table}[ht]
    \centering
    \begin{tabular}{p{4.5cm}p{1cm}p{1cm}p{1cm}p{1cm}p{1cm}p{1cm}}\hline
\textbf{Covariate} & 
\multicolumn{2}{c}{\textbf{\% at cohort entry}} & 
\multicolumn{2}{c}{\textbf{\% of value changed}} & 
\multicolumn{2}{c}{\textbf{\% mean}}\\
\hline\hline
Age ($\geqslant 75$) & 67 & 82& 0 & 0 & 66 & 82\\
Sex(female/male)& 54 & 52 & 0 & 0 & 54 & 52\\
\multicolumn{7}{c}{\textbf{Comorbidities/Medical score}}\\
$CHA_{2}DS_{2}VAS_{c}$  ($\geqslant 3$)& 80 & 89 & 0 & 0 & 79 & 88\\
Diabetes&34	&39	&0	&0&	34&	39\\
COPD/asthma&35&	51&	0&	0&	35&	49\\
Hypertension&81	&84	&0&	0	&81&	84\\
Malignant cancer&23	&36&	0&	0&	23&	36\\
Stroke&19	&16	&3&	5	&21	&18\\
Chronic kidney disease&33&	53&	6&	14&	36&	58\\
Heart disease  &66&	80&	7&	9&	70&	83\\
Major bleeding&28&	38&	9&	18&	33&	45\\
\multicolumn{7}{c}{\textbf{OAC use}}\\
DOAC&61&	51&	58&	47&	54&	38\\
Apixaban&31&	29&	26&	25&	27&	22\\
Dabigatran&11&	7&	13&	8&	10&	5\\
OAC&100&	100&	91&	94&	83&	70\\
High-dose-DOAC&39&	23&	41&	23&	34&	17\\
\multicolumn{7}{c}{\textbf{Concomitant medication use}}\\
Antiplatelets&52&	59&	43&	34&	24&	37\\
NSAIDs&7&	5&	11&	6& 3&	3\\
Statin&54&	52&	11&	11&	53&	49\\
Beta-Blockers&65&	63&	18&	12&	62&	62\\
\multicolumn{7}{c}{\textbf{Potential drug-drug interaction}}\\
DOAC: Antiplatelets&27&	25&	30&	27&	10&	11\\
DOAC: NSAIDs&4&	3	&7&	4&	2&	1\\
DOAC: Statin&30&	23&	33&	24&	29&	18\\
DOAC: Beta-Blockers&38&	30&	40&	32&	34&	22\\
\hline\\
    \end{tabular}
    \caption{Summary statistics of the baseline and time-dependent covariates}
    \label{A:tab:Tableone}
\end{table}
\section{Analysis using the time-dependent Cox model}\label{A:Cox}
Table \ref{A:tab:Cox} provides the crude (univariate) and adjusted hazard ratios from simple and multivariate time-dependent Cox models for death, respectively, and 95\% confidence intervals using the covariates in the analysis.
\begin{table}[ht]
    \centering
    \begin{tabular}{ccccc}\hline
\textbf{Covariate} & 
\multicolumn{2}{c}{\textbf{Crude HR}}&
\multicolumn{2}{c}{\textbf{Adjusted HR}}\\
 & 
\textbf{Estimate}&
\textbf{CI}&
\textbf{Estimate}&
\textbf{CI}\\
\hline\hline
Age ($\geqslant 75$) &2.25&	(2.09, 2.44)&	1.89&	(1.73, 2.06)\\
Sex(female/male)&0.92&	(0.86, 0.97)&	0.98&	(0.92, 1.04)\\
\multicolumn{5}{c}{\textbf{Comorbidities/Medical score}}\\
CHA2DS2VASc ($\geqslant 3$) &2.00&	(1.82, 2.20)	&1.01&	(0.90, 1.14)\\
Diabetes&1.23&	(1.16, 1.31)&	1.08	&(1.02, 1.15)\\
COPD/asthma&1.84&	(1.74, 1.96)&	1.49&	(1.40, 1.58)\\
Hypertension&1.20&	(1.10, 1.30)	&0.91&	(0.83, 0.99)\\
Malignant cancer&1.79&	(1.68, 1.90)&	1.45&	(1.36, 1.54)\\
Stroke&1.07 &	(1.00, 1.15)&	1.00	&(0.92, 1.07)\\
Chronic kidney disease&3.50	&(3.29, 3.73)&	2.10	&(1.96, 2.25)\\
Heart disease  &3.42&	(3.11, 3.76)&	2.41	&(2.18, 2.67)\\
Major bleeding&2.55&	(2.40, 2.70)&	1.38&	(1.29, 1.47)\\
\multicolumn{5}{c}{\textbf{OAC use}}\\
DOAC &0.06&	(0.06, 0.07)	&1.60	&(1.16, 2.21)\\
Apixaban&0.11	&(0.09, 0.13)	&0.88&	(0.68, 1.14)\\
Dabigatran&0.09&	(0.07, 0.12)&	0.77&	(0.53, 1.13)\\
OAC&0.03&	(0.02, 0.03)&	0.84	&       (0.66, 1.06)\\
High-dose-DOAC&0.07&	(0.06, 0.08)&	0.80&	(0.63, 1.01)\\
\multicolumn{5}{c}{\textbf{Concomitant medication use}}\\
Antiplatelets&1.37&	(1.28, 1.47)&	0.80&	(0.74, 0.86)\\
NSAIDs&1.14	&(0.97, 1.33)&	1.50&	(1.27, 1.76)\\
Statin&0.70	&(0.66, 0.75)&	0.82	&(0.76, 0.87)\\
Beta-Blockers&0.92	&(0.87, 0.98)&	1.37&	(1.28, 1.47)\\
\multicolumn{5}{c}{\textbf{Potential drug-drug interaction}}\\
DOAC: Antiplatelets&0.11&	(0.09, 0.15)&	1.09	&(0.81, 1.47)\\
DOAC: NSAIDs&0.15&	(0.09, 0.26)&	0.99&	(0.56, 1.78)\\
DOAC: Statin&0.08	&(0.07, 0.09)&	0.90&	(0.70, 1.15)\\
DOAC: Beta-Blockers&0.08&	(0.09, 0.10)&	0.60	&(0.47, 0.77)\\
\hline\\
    \end{tabular}
    \caption{Crude (univariate) and adjusted hazard ratios from simple and multivariate time-dependent Cox models for death, respectively, and 95\% confidence intervals using the covariates in the analysis.}
    \label{A:tab:Cox}
\end{table}
\section{Rational of the selection rules}
\label{A:ApplicationSelectionRuleExplain}
Selection rules 1 and 2: rule 1 is needed since when DOAC is in the model, and if Apixaban is selected, then the interpretation of the coefficient of Apixaban is the contrast of Apixaban and Rivaroxaban. If High-dose-DOAC is also in the model, the interpretation would be the contrast (e.g. log hazard ratio) of low-dose-Apixaban versus low-dose Rivaroxaban. However, without DOAC, the coefficient of Apixaban would represent a contrast against warfarin and Rivaroxaban combined, which is less interpretable. The same rationale applies to Dabigatran in rule 2.\\
Selection rule 3: it is needed because when OAC is in the mode, and if DOAC is selected, then the interpretation of the coefficient of DOAC is the contrast of DOAC and warfarin. However, without OAC, the coefficient of DOAC would represent a contrast against DOAC and warfarin or taking none of OAC combined, which is less interpretable. \\
Selection rule 4: it is needed since when DOAC is in the model, and if High-dose-DOAC is selected, then the interpretation of the coefficient of High-dose-DOAC is the contrast of high-dose-DOAC versus low-dose-DOAC, which is of interest. If Apixaban and Dabigatran are also in the model, the coefficient of High-dose-DOAC represents the contrast between high-dose-Rivaroxaban versus low-dose-Rivaroxaban. However, without DOAC in the model, these relevant interpretations would be lost. \\
\section{Grouping structure of the data analysis}
\label{A:GroupingStructure}
\begin{figure}
    \centering
    \includegraphics[width=0.6\linewidth]{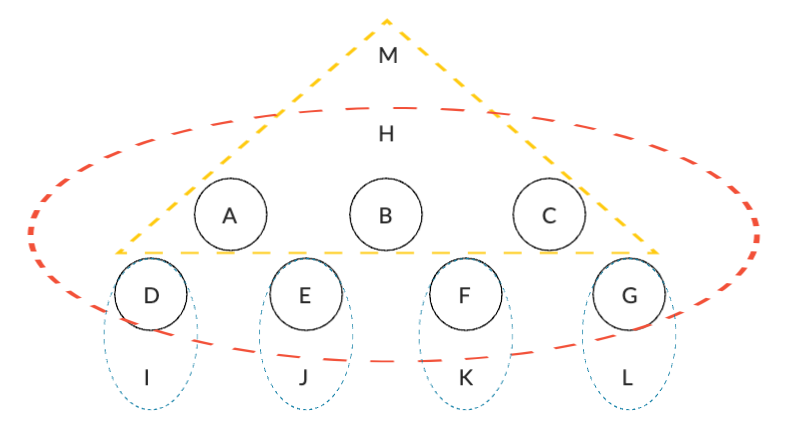}
    \caption{Grouping structure plot for the 13 groups}
    \label{fig:GrpStr}
\end{figure}

In our method, we need to specify the grouping structure to respect the selection rules. Thirteen variables are included in the 8 selection rules. For the convenience of grouping structure specification, we denote the 13 variables as such:\\
A: Apixaban, 
B: Dabigatran,
C: High-dose-DOAC,
D: DOAC: Antiplatelets,
E: DOAC: NSAIDs,
F: DOAC: Statin,
G: DOAC: Beta-Blockers,
H: DOAC,
I: Antiplatelets,
J: NSAIDs,
K: Statin,
L: Beta-Blockers,
M: OAC.\\
According to \citep{wang2021general, wang2022structured}. The grouping structure that relevant to these 13 variables should be
\begin{align*}
    \mathbb{g}_1=\{A\},
    \mathbb{g}_2=\{B\},
    \mathbb{g}_3=\{C\},
    \mathbb{g}_4=\{D\},
    \mathbb{g}_5=\{E\},
    \mathbb{g}_6=\{F\},
    \mathbb{g}_7=\{G\},
    \mathbb{g}_8=\{D, I\},\\
    \mathbb{g}_9=\{E, J\},
    \mathbb{g}_{10}=\{F, K\},
    \mathbb{g}_{11}=\{G, L\},
    \mathbb{g}_{12}=\{A-H\},
    \mathbb{g}_{13}=\{A, B, C, H, M\},
\end{align*}
13 groups in total. For the remaining 11 variables, each of them has one individual group. Therefore, we have 24 groups in total. 
For the 13 groups, we plot the grouping structure in Figure \ref{fig:GrpStr}. We can see that it presents a graph structure. Multiple groups are overlapped with and nested in other groups.  

To encode the grouping structure into our developed software, we need to specify two matrices, both of which are 24 by 24 matrices. The first matrix has 1 in the positions (4, 8), (5, 9), (6, 10), (7, 11), (1, 12), (2, 12), (3, 12), (4, 12), (5, 12), (6, 12), (7, 12), (1, 13) (2, 13), (3, 13), the rest are 0. The second matrix has 1 in the positions $(i, i), i=1,\dots,7, 13,\dots,24$, (9, 8), (10, 9), (11, 10), (12, 11), (8, 12), (8, 13), the rest are 0. For details, please see the help file in the R package.
\section{Visualization of the analysis}
\label{A:ApplicationVisual}
\begin{figure}
    \centering
    \includegraphics[width=0.7\linewidth,]{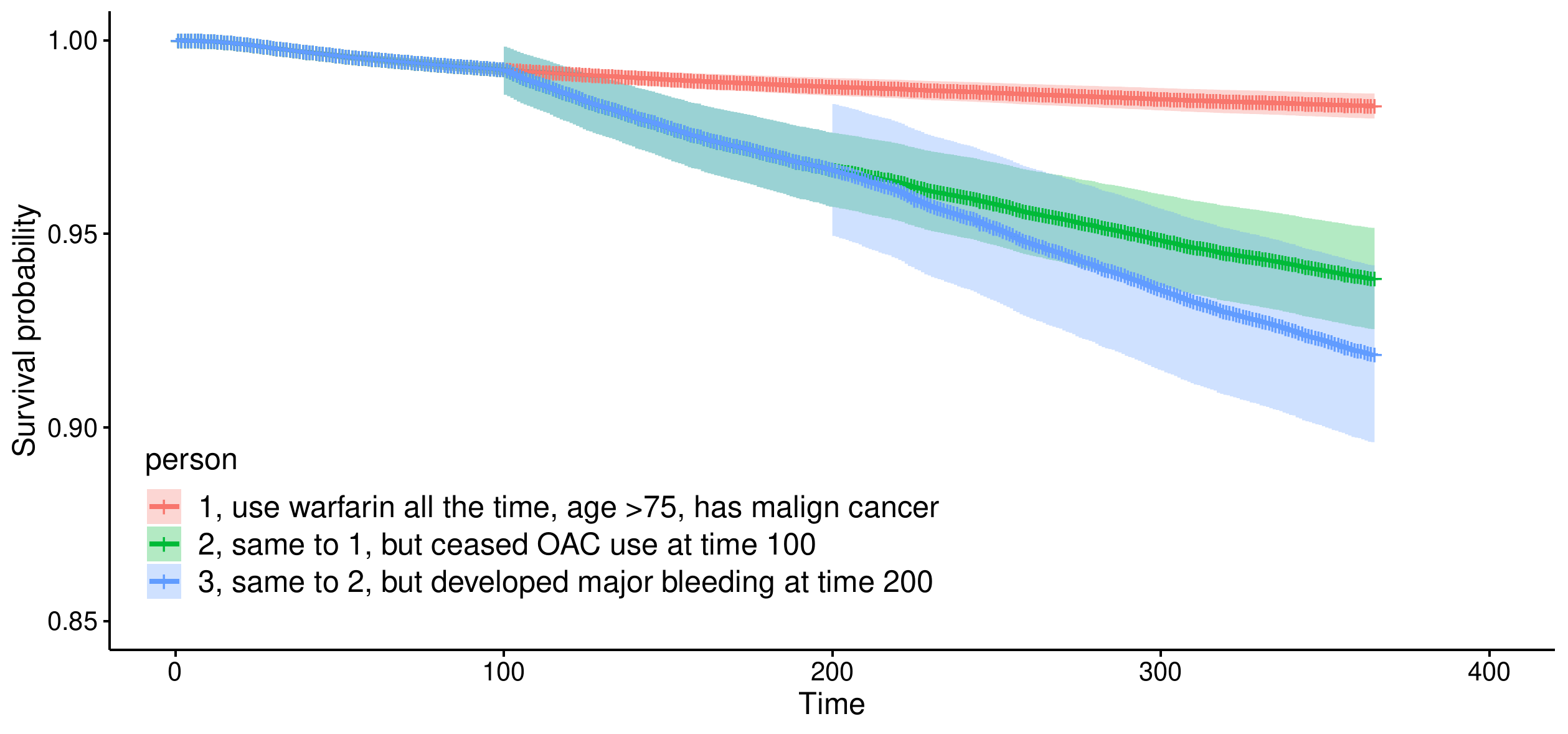}
    \caption{Estimated survival curves of three typical persons.}
    \label{fig:survivalplot}
\end{figure}

We artificially create three hypothetical patients' disease progression. Suppose person 1, age $\geqslant 75$, who only used the drug warfarin, had only malign cancer among all the disease variables included in the data. Person 2 has the same profile as person 1 except they ceased warfarin at time 100 during the follow-up, while other statuses stayed the same. Person 3 developed major bleeding at time 200, and all other statuses were the same as person 2. Figure \ref{fig:survivalplot} shows the survival curves of the three people, estimated by the time-dependent Cox model using the covariates that were selected by our method. 
As we can see, person 1 (in red) has the highest estimated survival probability. The survival probability of person 2 drops immediately after the cease of warfarin. Similarly, the survival probability drops significantly at time 200 due to the major bleeding. Note that Figure \ref{fig:survivalplot} only intends to show, as an example, how the survival probability can vary according to time-dependent covariates, rather than predicting the survival probability of the hypothetical cases. All the covariates involved in the study are internal covariates that relate to the outcome in the sense that the covariates can be measured only among the patients who are still at risk of the event (alive). 
\bibliographystyle{NJDnatbib}
\bibliography{ama/refs}
\end{document}